\documentclass[bibyear]{aa}\usepackage{txfonts}% if the references are not structured 
\usepackage{xcolor}
\usepackage{graphicx}
\usepackage{natbib}
\usepackage{txfonts}
\usepackage{lipsum}
\usepackage{longtable} % for 'longtable' environment
\usepackage{lscape} % for 'landscape' environment
\usepackage[hidelinks]{hyperref}
\hypersetup{colorlinks=true,linkcolor=blue,citecolor=blue,urlcolor=blue}
\usepackage{placeins}
\begin{document}

   \title{Re-analysis of 10 Hot-Jupiter Atmospheres with disequilibrium chemistry retrieval}

   %\subtitle{I. Overviewing the $\kappa$-mechanism}

   \author{Deborah Bardet\inst{1}
        \and
            Quentin Changeat\inst{2,3}
        \and
            Olivia Venot\inst{1}
        \and
            Emilie Panek\inst{4}
            }

   \institute{Université Paris Cité and Univ Paris Est Creteil, CNRS, LISA, F-75013 Paris, France \\
        \email{deborah.bardet@lmd.ipsl.fr}
        \and
            Kapteyn Institute, University of Groningen, 
            9747 AD Groningen, NL
        \and
            Department of Physics and Astronomy, University College London, London, UK
        \and
            Institut d’Astrophysique de Paris (CNRS, Sorbonne Université), 98bis Bd Arago, 75014 Paris, France
    }

   \date{Received ***; accepted ***}

% \abstract{}{}{}{}{} 
% 5 {} token are mandatory
 
  \abstract
    {Constraining the chemical structure of exoplanetary atmospheres is pivotal for interpreting spectroscopic data and understanding planetary evolution. Traditional retrieval methods often assume thermochemical equilibrium or free profiles, which may fail to capture disequilibrium processes like photodissociation and vertical mixing. This study leverages the TauREx 3.1 retrieval framework coupled with FRECKLL, a disequilibrium chemistry model, to address these challenges.}  
   {The study aims to (1) assess the impact of disequilibrium chemistry on constraining metallicity and C/O ratios; (2) evaluate the role of refractory species (TiO and VO) in spectral retrievals; (3) explore consistency between transit and eclipse observations for temperature and chemical profiles; and (4) determine the effects of retrieval priors and data reduction methods.}
   {Ten hot-Jupiter atmospheres were reanalyzed using Hubble Space Telescope (HST) WFC3 data in eclipse and transit. The TauREx-FRECKLL model incorporated disequilibrium chemistry calculations with a Bayesian framework to infer atmospheric properties. Retrieval scenarios included tests with and without TiO/VO and comparisons across different data reduction pipelines.}
   {The disequilibrium approach significantly altered retrieved metallicity and C/O ratios compared to equilibrium models, impacting planet formation insights. TiO/VO additions improved fits for only two planets, with limited effect on parameter convergence. Retrievals reconciled transit and eclipse temperature profiles in deeper atmospheric layers but not in upper layers. Results were highly dependent on spectral resolution and retrieval priors, emphasizing limitations of HST data and the need for broader spectral coverage from instruments like JWST.}
   {This study demonstrates the feasibility and importance of incorporating disequilibrium chemistry in atmospheric retrievals, highlighting its potential for advancing our understanding of exoplanetary atmospheres with next-generation telescopes.}


   \maketitle
%
%-------------------------------------------------------------------

\section{Introduction}
Correctly constraining the chemical structure of exoplanetary atmospheres is a crucial element in the inversion of spectroscopic data, since it provides important information about their physical and climatic characteristics, and above all about their evolutionary history \citep{Ober:11,Madh:16,Mord:16,Brew:17,Eist:18,Turr:21,Loth:21}.

The chemical structure of an exo-atmosphere was originally recovered using either profiles constant with altitude (a single free parameter representing each molecule) or assuming thermochemical equilibrium \citep{Whit:58, Madh:09,Lee:12_NEMESIS,Line:13,Wald:15, AlRe:21}, which requires computing the chemistry state by minimizing the Gibbs free energy of the system. 
The latter is advantageous in terms of degrees of freedom, since it only requires two free parameters for metallicity and the carbon-to-oxygen (C/O) ratio chosen for their natural links to planetary formation and evolution processes.
However, assuming thermochemical equilibrium is a strong assumption, especially considering the orbital geometry of exoplanets: while being close to their host stars implies very high atmospheric temperatures---which promote rapid chemical kinetics and therefore chemical equilibrium---other processes that could drive the atmosphere out-of-equilibrium are also enhanced. 
For instance, the intense irradiation in the Ultraviolet (UV) range triggers important processes of photodissociation, processes driving the chemical composition away from chemical equilibrium (and thus missing in equilibrium calculations). 
Examples of planetary atmospheres in our solar system and recent JWST results \citep{Tsai:23SO2, Dyre:24} suggest that exo-atmospheres are incompatible with this thermochemical equilibrium assumption.
Since some attempts have been done to explore other and more complex chemical assumptions in exo-atmosphere retrieval, such as two-layer profiles chemistry \citep{Chan:19}, a chemical relaxation disequilibrium \citep{Kawa:21_diseqchem}, or an equilibrium offset \citep{Madh:09}. 
Nonetheless, all of those chemical assumptions are approximate solutions of chemical disequilibrium and do not directly handle the kinetic reactions. 

The need to account for disequilibrium processes (vertical mixing including quenching and photodissociation) in spectral data inversion calculations is all the more obvious after the latest observations of SO$_2$, produced by processes initiated by photochemistry \citep{Tsai:23SO2,Dyre:24}, and the lack of photochemistry processes in data retrieval calculations remains one of the main sources of error. 
To reinforce this argument, simulations using kinetics calculations (i.e., taking into account disequilibrium processes such as mixing and photochemistry) have proven that chemical equilibrium is inadequate in many scenarios \citep[e.g.,][]{Mose:11,Mose:13,Mose:16,Veno:12,Veno:14,Veno:20,Veno:20_WASP43b,Morl:17,Mola:19,Moll:20,Kawa:21_diseqchem,Tsai:21}. 
To ensure the unbiased interpretation of exoplanet observations, it is crucial to properly represent the chemical processes within those atmospheres, in particular with the modern and future telescopes, whose resolution enables more precise and complete physical and chemical characterisation -- see the first analyses of James Webb Space Telescope data \citep{JWST:23,Tsai:23SO2,Dyre:24}.

For this purpose, the retrieval model TauREx 3.1 \citep{AlRe:21} has been coupled with FRECKLL, the disequilibrium chemical kinetic Python model, thanks to the plugin functionality \citep{AlRe:24}. 
In this study, the authors have presented this new coupled model, as well as retrieval test cases on simulated HD 189733~b observations at JWST resolution to demonstrate the viability of total disequilibrium chemistry retrievals and the ability for JWST to detect disequilibrium processes. 

The aim of the present paper is to address some of the exoplanet and observation analysis questions raised by population studies \citep{Sin:16_cloudyhotJup, Tsia:18, Min:20, Estr:22, Chan:22_5keyQuestions, Edwa:23, Saba:24}, using this coupled TauREx-FRECKLL model for the first time on real observations. 
We consider 10 Hot-Jupiter spectra, from the HST-WFC3 G141 grism, in both eclipse and transit. 
To that end, we are considering the following questions: 
(1) Does the addition of a kinetic chemical model to the retrieval algorithm make it possible to better constrain metallicity and the C/O ratio, making them viable observables for a better understanding of exoplanet formation? \cite{Chan:22_5keyQuestions} have pointed out that, using either a free or a thermochemical equilibrium hypothesis for their retrieval calculations, the C/O ratio, as well as metallicity are difficult to constrain in HST observations. However, those two are crucial to determine precisely the planetary formation. If Hot Jupiters are actually formed via a three-step process \citep{Mizu:80,Bode:86,Ikom:00} -- namely solid core accretion, runaway gas accretion, and migration --, the current composition of the giant exoplanets must have remained the same (or very close to it) as in the early stages of planetary formation, i.e. substellar in heavy elements such as C, O, and refractory elements because most of the heavy elements would be sequestered in the cores. Hence, we are attempting to better constrain the C/O ratio and metallicity by taking into account kinetic chemistry calculations in HST retrievals.    

(2) How important are the chemical species TiO and VO in the exoplanet atmospheric spectra?  
Due to the fact that HST instruments are not sensitive to N- and S-bearing molecules, the detection, and afterwards the analysis of elemental ratios such as N/O and S/O \citep{Turr:21} have been unexplored with HST data and could not help to the determination of planet formation scenarios. 
Another way to infer planetary formation is resorting to refractory elements analysis \citep{Loth:21}. 
Hence, previous observational studies have turned to the analysis of refractory elements as TiO, VO and FeH, which have been detected in HST eclipse spectra. 
For instance, \cite{Chan:22_5keyQuestions} have shown that, in a population study encompassing 25 hot gaseous planets viewed in eclipse, the hottest planets in their sample (T $>$ 2000 K) have inverted thermal profiles with signatures from thermal dissociation (H$^-$) and refractory species (TiO, VO, or FeH), and the spectra in the HST wavelength range are not consistent with a simple blackbody emission. 
However, the authors advertised that equilibrium chemistry retrievals are not the preferred solution for a large number of planets in their selected population, and strongly suggest that disequilibrium mechanisms might be important.
Therefore, considering disequilibrium hypothesis in retrieval calculation of our set of planets, we will explore the impact of the addition of TiO and VO in improving spectra inversion.

(3) Does the use of a kinetic chemistry model coupled with the retrieval model allow consistent eclipse and transit retrieval for individual exoplanet, in term of temperature and chemistry structures? Transit spectra are primary sensible to the atmospheric scale height of the planet; depending of planet radius, planet mass, and the mean temperature. In transit view, the mean temperature does not take into account the possible inhomogeneities between morning and evening terminator compositions, implying large degenerencies for retrieval calculations (for both temperature and chemical structures), due to the geometry of the observed spectra. Hence, the temperatures of exoplanets inverted from transit data and reported in literature are biased from several hundred degrees for the better, to more than thousand degrees for the worst. This systematic bias has been obtained with 1D retrieval codes describing an approximate chemical composition -- either following a free or a thermochemical equilibrium assumption --, adding further significant biases in the retrieved chemical abundances and so retrieved temperature structure.  To compensate for this bias in one-dimensional models, it is essential to turn to 2D or 3D inversion calculations to take into account the inhomogeneities of morning and evening limbs \citep{MacD20}. Bi- or tri-dimensional retrieval calculations, with disequilibrium chemistry assumption is out of the scope of the present paper. However, the coupling between TauREx 3.1 and FRECKLL permits to verify the consistency between eclipse and transit inverted spectra using a 1D retrieval model. This point aims to draw conclusions about the level of information about the temperature structure that could be extracted from HST observations in transit, by considering a kinetic chemical scheme in retrieval calculations.
    
The paper is structured as follows: Section \ref{sec:methodo} presents the data processing, retrieval procedures, and opacity materials as well; 
results of retrieval calculations are found in Section \ref{sec:results}; and
finally, discussions and concluding remarks are drawn in Section \ref{sec:discussion}.

\section{Methodology}
\label{sec:methodo}
\subsection{Data and processing procedure}

Our study encompasses data for 10 Hot-Jupiters observed in eclipse with the HST-WFC3 G141 grism: HAT-P-2 b \citep{Salz:15}, HD 189733~b \citep{Addi:19}, HD 209458~b \citep{Bars:17}, Kepler-13 A b \citep{Bata:13}, TrES-3 b \citep{Sout:11}, WASP-4 b \citep{Huit:17}, WASP-19 b \citep{CortZula:20}, WASP-43 b \citep{Espo:17}, WASP-74 b \citep{Manc:19}, and WASP-77 A b \citep{CortZula:20}; as well as data in transit with the HST-WFC3 G141 grism for four of those planets: HD 189733~b, HD 209458~b, WASP-43 b, and WASP-74 b. 
For our retrieval calculations, we use the consistent set of reduced data from Iraclis \citep{Tsia:16_Iraclis}, obtained in eclipse from \citet{Chan:22_5keyQuestions} and in transit from \citet{Tsia:16_Iraclis} and \citet{Edwa:23}.

All of the HST-WFC3 data used in those three studies were processed following the same procedure to ensure consistency between planets \citep[detailed in][]{Tsia:16_Iraclis,Chan:22_5keyQuestions,Edwa:23}.
We summarise this procedure in what followed.
\cite{Tsia:16_Iraclis,Chan:22_5keyQuestions,Edwa:23} used the Iraclis pipeline, a highly specialised software for processing WFC3 spatially scanned spectroscopic images \citep{Tsia:16_Iraclis}. 
All data have been reduced following an eight-step-process, namely: zero-read subtraction, reference-pixel correction, non-linearity correction, dark current subtraction, gain conversion, sky background subtraction, calibration, flat-field correction, and bad-pixel/cosmic ray correction. 
Then, the white (1.088--1.68$\mu$m) and the spectral light curves were extracted from the resulting reduced images, taking into account the geometric distortions caused by the tilted detector of the WFC3 IR channel.

The white light curves for each planet (either in eclipse or transit data) were fitted using the transit model package {PyLightcurve} \citep{Tsia:16_pylightcurve}, only allowing the planet-to-star radius ratio and the midtransit (mideclipse) time as free parameters for the transit (eclipse) data. 
Time dependent systematics, the long-term and short-term ``ramps'', affect each HST visit and each HST orbit, respectively. 
The first ramp has been corrected using a linear behaviour formula, and the second has been corrected using an exponential behaviour. 
The white light curves fitting included this ``ramps'' correction, as well as the uncertainties per pixel, as propagated through the data reduction process with Iraclis. 
Additional scatter not explained by the ramp model has been corrected as well, with a scaling of the uncertainties in the individual data points, for their median to match the standard deviation of the residuals, and repeated the fitting  again with the planet-to-star radius ratio and the midtransit (mideclipse) time as free parameters for the transit (eclipse) data \citep{Tsia:18}. 

The spectral light curves were fitted to a transit model (with the planet-to-star radius ratio being the only free parameter), considering a model for the systematics that includes the white light curve \citep[divide-white method,][]{Krei:14_cloudGJ1214b} and a wavelength-dependent, visit-long slope \citep{Tsia:16_Iraclis}. 
In the same way as for the white light curve, an initial fit has been performed using the pipeline uncertainties, and then refitted while scaling these uncertainties for their median to match the standard deviation of the residuals. 

\subsection{Retrieval procedure}

\begin{table*}
    \caption[]{List of the free parameters and their uniform priors in the retrievals.}
    $$
    \label{tab:retrieval_settings} 
    \begin{tabular}{llll}
        \hline
                Type      &  Parameter           & Mode    &   Prior  \\
                \hline
        \hline
                Transmission    & Planet~radius ${(\mathrm{R_{planet}})}$  & linear  & 0.5 : 5.0 ${[\mathrm{R_{J}}]}$  \\
                        & Temperature points $i$ ${(\mathrm{T_i})}$  & linear &  500 : 3900 ${[\mathrm{K}]}$ \\
                        & C/O ratio                     & linear  & 0.01 : 1.1 \\
                        & Metallicity ${(\mathrm{Z})}$  & log  & $10^{-2} : 10^{3}$  ${(\mathrm{to~solar})}$\\
                                                & Eddy diffusion coefficient ${(\mathrm{K_{zz}})}$    & log  & $10^{6} : 10^{14}$ ${[\mathrm{cm^{2}~ s^{-1}}]}$ \\
        \hline
        \hline
                Emission    & Temperature points $i$ ${(\mathrm{T_i})}$  & linear &  500 : 3900 ${[\mathrm{K}]}$ \\
                    & C/O ratio                     & linear  & 0.01 : 1.1 \\
                    & Metallicity ${(\mathrm{Z})}$  & log  & $10^{-2} : 10^{3}$  ${(\mathrm{to~solar})}$\\
                    & Eddy diffusion coefficient ${(\mathrm{K_{zz}})}$    & log  & $10^{6} : 10^{14}$  ${[\mathrm{cm^{2}~s^{-1}}]}$\\
                    & Abundances (free: TiO, VO)    & log  & $10^{-12} : 10^{2}$  ${(\mathrm{VMR})}$\\ 
                \hline
        \hline
    \end{tabular}
    $$
\end{table*}

Our systematic retrieval calculations are computed using the TauREx 3.1 atmospheric retrieval code \citep{AlRe:21}, and the retrieval parameters used for this study are summarised in Table \ref{tab:retrieval_settings}. 
The modelled planetary atmospheres assume the plane-parallel assumption, hence atmospheric column is uniformly discretized in log-pressure with 20 layers per decade, for 80 levels from 10$^{1}$ to 10$^{-3}$~bar.
The temperature profile is determined using a heuristic N-point profile: discretized by 5 temperature points, one for each decade of pressure.
Each of these 5 temperature points is assigned a prior of 500-3900~K, but their pressure level is fixed \citep[][a sensitivity study on the number of temperature points to retrieve the temperature profile is available in Appendix \ref{app:sensitivity_npoints_Tp}]{Chan:21_phasecurve_WASP43b}.
The C/O ratio had a uniform a priori of 0.01-1.1, and the metallicity $Z$ had a logarithmic a priori of $10^{-2}-10^{3}$ (to solar).
Here, the metallicity and the C/O ratio are both varying, with the metallicity controlling all the elements heavier than He and the C/O ratio varying from the fiducial solar elemental abundance with the abundance of carbon. 
Hence, the abundance of oxygen is fixed at the value recommended by \cite{Lodd:10}, (i.e., log(He) = 10.925, log(C) = 8.39, log(O) = 8.73, and log(N)= 7.86).
For transmission spectra, the planetary radius R$_{planet}$, at the reference pressure of 10 bar, has a uniform a priori of 0.5-5.0 R$_J$. 
The parameter space is explored using the MultiNest nested sampling algorithm \citep{Fero:08,Fero:09,Fero:19}, implemented by the PyMultiNest python wrapper \citep{Buch:14}, with 400 live points and a log-likelihood tolerance of 0.5. 
Because we use the nested sampling algorithm MultiNest, the computation of the Bayesian evidence for each model, here denoted E, is automatic. 
Then, the difference in ln(E) between two models M1 and M2 can be used for model selection and to compare the ability of the two models to explain the observed spectra \citep{Kass:95_BayesFactor,Tsia:16_Iraclis}. 
Therefore, we use in what follows the Bayesian evidence to compare the results with the ``free runs'' (assuming constant abundances as a function of
altitude of the considered molecular species) from \citet{Chan:22_5keyQuestions} (for eclipse data) and \citet{Edwa:23} (for transit data).

The disequilibrium chemical calculations are done by the FRECKLL model, using the Python plugin developed by \citet{AlRe:24}. 
Contrary to thermochemical equilibrium models, predicting the chemical state of a planet's atmosphere by minimizing the Gibbs free energy of the system, chemical kinetic models integrate a system of differential equations (formed by continuity equations of each species considered, describing the temporal evolution of their abundance) until a steady state is reached.
In FRECKLL, continuity equations of each species evolve the stiff ordinary differential equation (ODE) solver VODE package \citep{Brow:89_VODE}, with the initial atmospheric state initialized at thermochemical equilibrium composition.
In addition to the metallicity and C/O ratio parameters, FRECKLL also needs the definition of the vertical eddy diffusion parameter K$_{zz}$, given as a constant value in the present study. 
For the validation study, we employ the full \cite{Veno:20} network, including 108 species, made of H, He, C, O, and N (with up to two carbon atoms i.e., C$_0$-C$_2$ chemical kinetic network), 1906 reactions, and 55 photodissociations (all the photolysis data are contained in the scheme, and the description of the UV spectrum used for the host star of each planet is detailed in Appendix \ref{app:stellar_flux}).

For each eclipse spectra, we have carried out several retrieval calculations to address the questions formulated in the Introduction: (i) a retrieval test with FRECKLL only (hereafter named FRECKLL-only), a test with FRECKLL plus TiO following the ``free'' chemical structure (i.e., a constant-with-altitude profile, hereafter named FRECKLL-TiO), (ii) a test with FRECKLL plus VO (``free'' chemical structure, hereafter named FRECKLL-VO), and (iii) FRECKLL plus both TiO and VO as well, named FRECKLL-TiO-VO. 
In addition, planets for which the FRECKLL-only retrieval result described a bi-modal metallicity distribution (HAT-P-2~b, HD 189733~b, WASP-19~b, and WASP-74~b, see Appendix \ref{app:indiv_eclipse}), we have carried out additional tests: we have separated the range of metallicity priors into two sub-ranges ($[10^{-2}:10^{1}]$ and  $[10^{1}:10^{3}]$  ${(\mathrm{to~solar})}$) and repeated the TiO and VO addition tests for each of these two sub-ranges. 
Those additional tests are motivated by the very nature of the Multinest sampling: when two solutions are found, the number of live points (representing at first order the resolution of the sampling, here set at 400) is halved to explore the two solutions as best as possible. 
Artificially separating the metallicity regimes with two distinct retrieval calculations enables us to maintain the number of live points needed to explore a single solution, and to maintain resolution consistency throughout the set of spectra. 
The Bayesian evidences are then compared to determine the most probable solution between the two modes we have forced through the separation of metallicity regimes. 

For transit spectra, only two retrieval tests have been done, both only with FRECKLL, using either an isothermal profile for the retrieved temperature calculations, or a 5-point thermal profile (see Appendix \ref{app:indiv_transit}).  

\subsection{Opacity sources}
For our re-analyse study, we have selected molecular line lists from the Exomol project \citep{Tenny:16,Chub:21}, HITEMP \citep{Roth:14}, and HITRAN \citep{Gord:16}. 
We draw the attention of the reader that, as the FRECKLL C$_0$-C$_2$ chemical network only includes chemical species made of H, He, C, O, and N, chemical species such as TiO and VO are not included in FRECKLL.  
Hence, our set of exoplanets has been selected to be valid for FRECKLL regime i.e., set of exoplanets without inversion in their thermal profile (from previous studies \citealt{Chan:22_5keyQuestions,Edwa:23}), without TiO and/or VO spectral signature, and with an equilibrium temperature under 2500 K. 
But in what follows we will show a case for which we can extend the retrieval assumptions and improve the results by including TiO and/or VO following the ``free'' chemistry assumption (WASP-77~A~b).
We have therefore selected the same molecular cross sections at resolution R = 15,000 as \cite{Chan:22_5keyQuestions}, namely: H$_2$O \citep{Bart:17,Poly:18}, CH$_4$ \citep{Hill:13,Yurc:14}, CO \citep{Li:15_COline}, CO$_2$ \citep{Yurc:20_CO2}, as well as HCN \citep{Barb:14_Exomol}, H$_2$CO \citep{AlRe:15}, CN \citep{Syme:21}, C$_2$H$_2$ \citep{Chub:20Exomol}, C$_2$H$_4$ \citep{Mant:18}, NH$_3$ \citep{AlDe:15,Cole:19}, TiO \citep{McKem:19}, VO \citep{McKem:16}.
The opacity sources for those latter (TiO and VO) are only used by the radiative transfer calculation into TauREx.
Atomic and ionic species were not included as they do not absorb in the wavelength range considered here.
We have also included collision-induced absorption (CIA) of the H$_2$–H$_2$ \citep{Abel:11,Flet:18HydrogenDimers} and H$_2$–He \citep{Abel:12} pairs as well as opacities induced by Rayleigh scattering \citep{Cox:15}.

\section{Results}
\label{sec:results}

\begin{figure}
   \centering
   \includegraphics[width=0.45\textwidth]{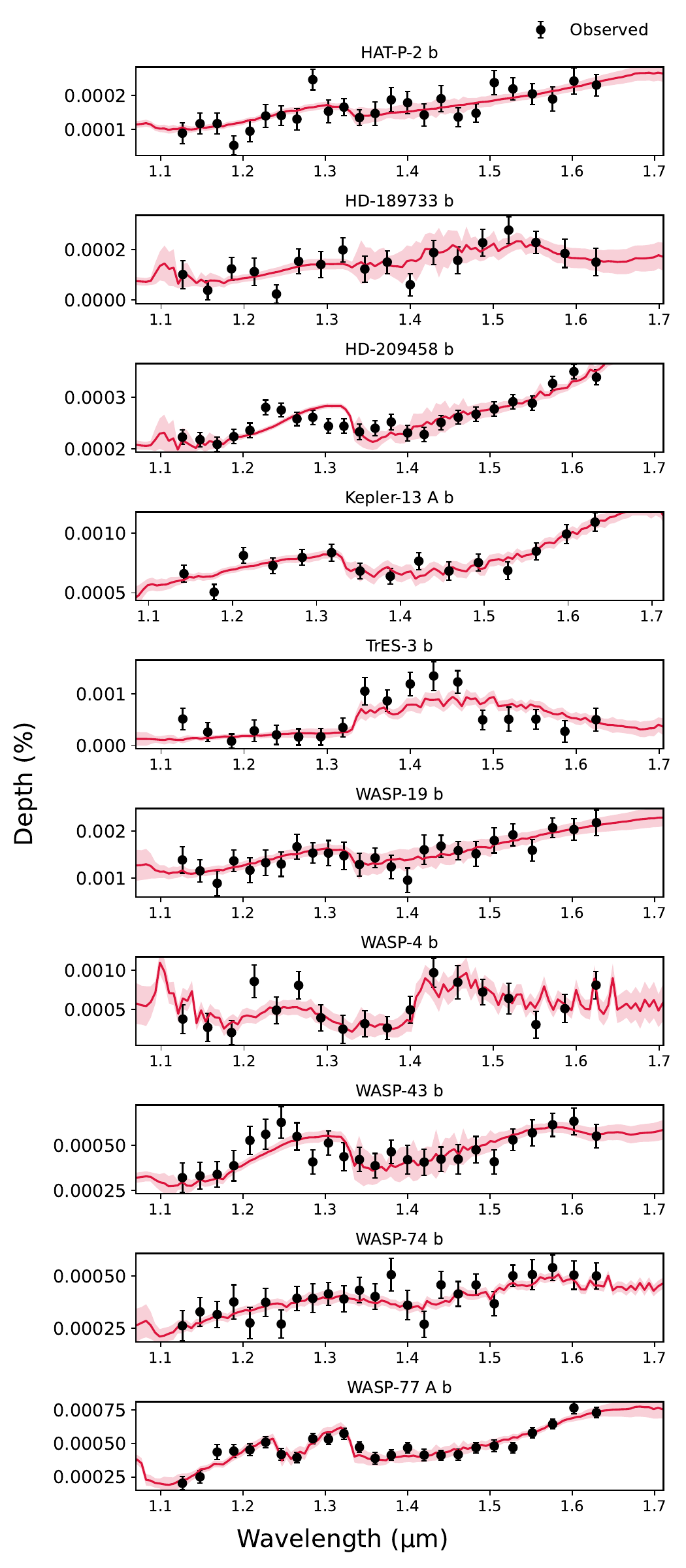}
   \caption{FRECKLL-only fit spectra of all the planets considered in this study -- except for HD 189733~b (FRECKLL-log(Z)$<$0, for which restricting the metallicity prior to the subsolar region improved ln(E) by 30 compared to the reference ``FRECKLL-only'' retrieval) and  WASP-77~A~b (FRECKLL-TiO retrieval, for which adding TiO really improve the fitting spectrum) -- for the HST observations in eclipse. Individual analyses and additional retrievals can be found in Appendix \ref{app:indiv_eclipse}.}
    \label{fig:emission_superposed_spectra}
\end{figure}

\begin{figure}
   \centering
   \includegraphics[width=0.45\textwidth]{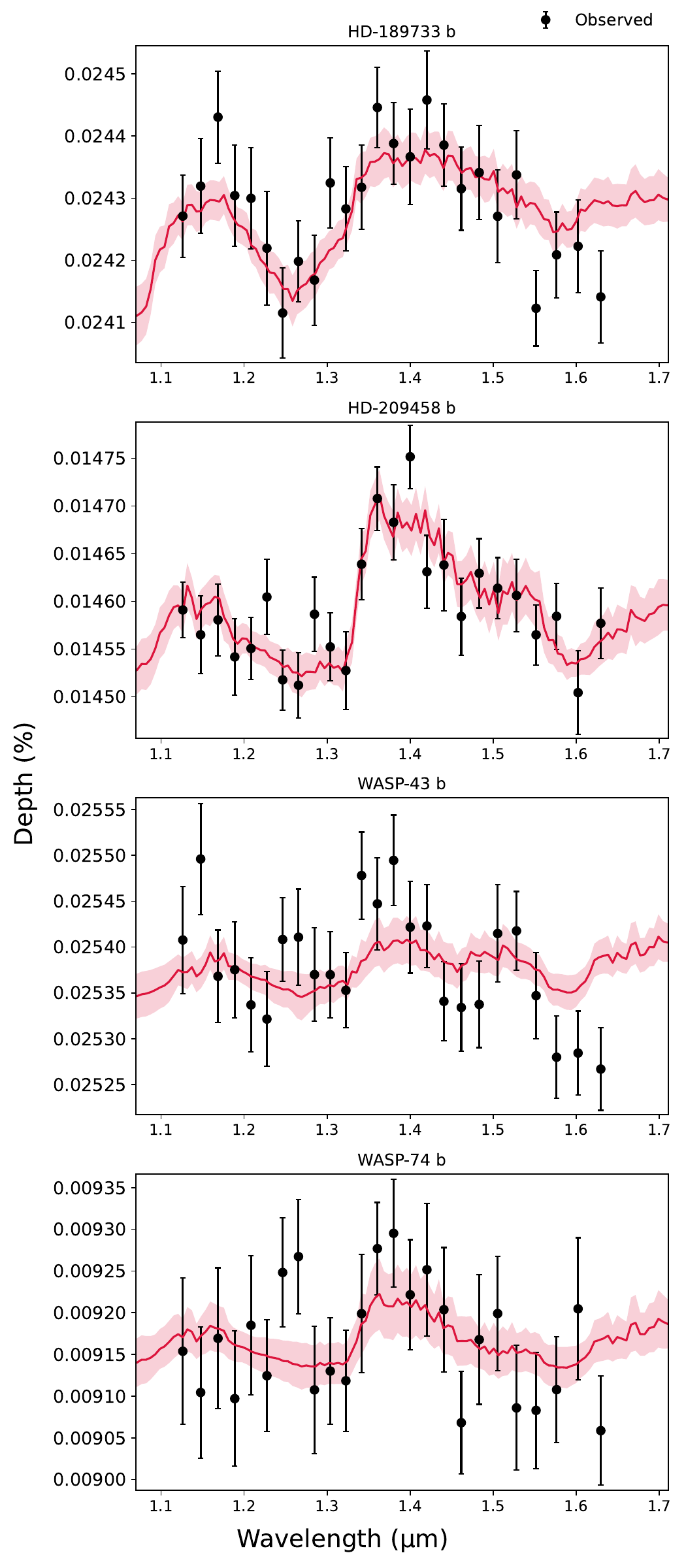}
   \caption{FRECKLL-isothermal fit spectra of the 4 planets considered in this study for the HST observations in transit. Individual analyses and additional retrievals can be fund in Appendix \ref{app:indiv_transit}.}
    \label{fig:transmission_superposed_spectra}
\end{figure}

\begin{figure*}
   \centering
   \includegraphics[width=0.95\textwidth]{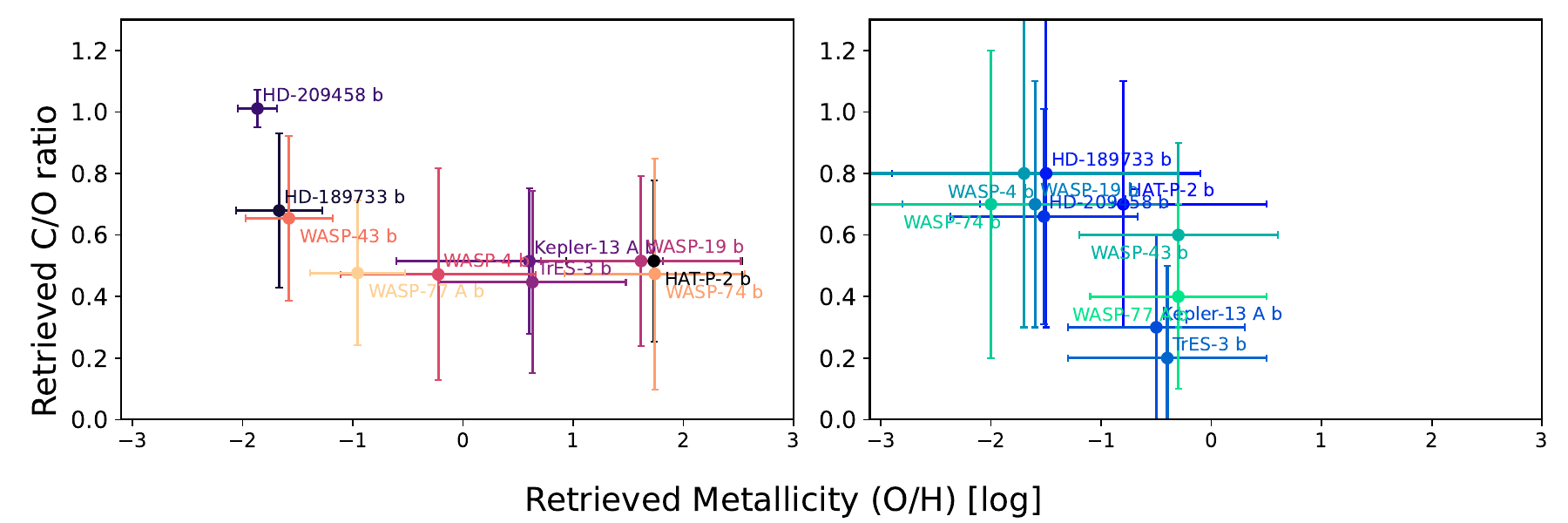}
   \caption{Eclipse retrievals: (Left panel) Metallicity (O/H) and C/O retrieved from the FRECKLL-only retrievals, except for HD 189733~b (FRECKLL-log(Z)$<$0, for which restricting the metallicity prior to the subsolar region improved ln(E) by 30 compared to the reference ``FRECKLL-only'' retrieval) and  WASP-77~A~b (FRECKLL-TiO retrieval, for which adding TiO really improves the fitted spectrum). C/O remains very difficult to retrieve because HST observations lack sensitivity to carbon-bearing species. 
   (Right panel) From \cite{Chan:22_5keyQuestions} (Table D1): Metallicity and C/O retrieved from HST free chemistry retrieval taking into account free chemistry profiles for H$_2$O, CH$_4$, CO, CO$_2$ for all planets, except WASP-77~A~b for which there are additional free profiles for TiO, VO, FeH and e$^-$. }
    \label{fig:emission_CO_Z}
\end{figure*}

The HST eclipse and transit observations retrieved with the coupled TauREx-FRECKLL model are shown in Figure \ref{fig:emission_superposed_spectra} and Figure \ref{fig:transmission_superposed_spectra}, respectively. 
We adopt a conservative approach and show the FRECKLL-only fits for eclipse (except for WASP-77 A b which is FRECKLL-TiO), and FRECKLL-isothermal fits for transit, unless another retrieval model is significantly preferred -- i.e., the difference in $ln(E)$ between two retrieval runs $\Delta ln(E) > 5$ (see Tables \ref{tab:eclipse_retrieval_results} and \ref{tab:transit_retrieval_results} for detailed retrieval results). 
Hence, for WASP-77~A~b, the addition of TiO as a ``free'' chemical species improved the retrieval fit, resulting of a $ln(E) > 199$ (for the ``FRECKLL-only'' case, $ln(E) = 195$).
Therefore, for the population diagnostics figures (Figs \ref{fig:emission_superposed_spectra}, \ref{fig:emission_CO_Z}), we will consider the FRECKLL-TiO retrieval results in eclipse for this planet.

\begin{figure}
   \centering
   \includegraphics[width=0.45\textwidth]{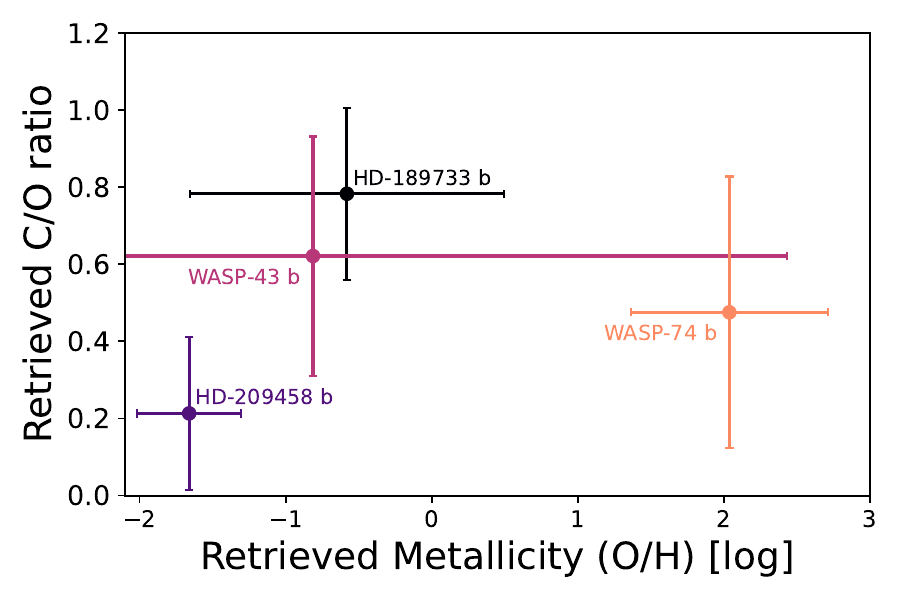}
   \caption{Transit retrievals: Metallicity (O/H) and C/O retrieved from the FRECKLL-isothermal retrievals. C/O remains very difficult to be retrieved because HST observations lack sensitivity to carbon-bearing species.}
    \label{fig:transmission_CO_Z}
\end{figure}

To infer metallicity and C/O ratio of exoplanets, population studies rely on the detection of water vapour, C-bearing species, or from chemical equilibrium calculations. 
\cite{Chan:22_5keyQuestions} (hereinafter C22) argued that, if detected, water vapour can be used as a proxy of metallicity, allowing the estimation of O/H ratio. 
While directly inferring the C/O ratio from C-bearing species remains difficult with HST data (as this telescope lacks sensitivity for this family of molecules), our set of planets for this population study depicts a clear H$_2$O spectral signature, which is known to also affect C/O ratio \citep{Rocc:16,Drum:19}, and an equilibrium temperature inferior to 2500~K (``cold'' hot Jupiters) preventing any degeneration of metallicity and C/O from dissociated water vapour (only Kepler-13 A b has an equilibrium temperature warmer than 2500~K to test the limit of our model).
\\

The use of a disequilibrium model to analyse HST data may have a fundamental impact on previous conclusions drawn using other chemistry assumptions. In particular, it could affect the retrieved metallicity and C/O ratio of exoplanets, and thus the reservoir for planet formation to which they belong.
This is shown in Figure \ref{fig:emission_CO_Z} and Figure \ref{fig:transmission_CO_Z} where we present the metallicity and C/O ratio inferred from our disequilibrium chemical model, respectively for eclipse and transit data.  
As evidenced in previous population studies \citep{Tsia:18,Chan:22_5keyQuestions,Edwa:23}, our results suggest two distinct groups of planets, which is particularly evident for the transit data. 
This overall picture is not consistent with C22, displaying all planets considered here in the negative range of metallicity.
Therefore, for individual planets, our results can differ significantly from results assuming free or chemical equilibrium, with a moderate impact on the Bayesian evidence for disequilibrium chemistry retrievals (results from Table D1 of C22 HST free chemistry retrieval taking into account free chemistry profiles for H$_2$O, CH$_4$, CO, CO$_2$, for all planets and in addition free profiles for TiO, VO, FeH, e$^-$ in the case of WASP-77 A b have been plotted on Figure \ref{fig:emission_CO_Z} right panel for easy comparison).  
For example, for the disequilibrium chemistry case, HD 189733~b has a subsolar metallicity (log(Z)$\sim$-1.5 in eclipse and $\sim$-0.5 in transit), with a C/O ratio between 0.7 (eclipse) and 0.8 (transit); consistent with the free chemistry retrieval carried by C22 which placed HD 189733~b as a subsolar metallicity planet (log(Z)$\sim$-1.5) as well, with a C/O ratio equal to 0.8. Similar results are obtained for HD 209458~b, except that in our case the C/O ratio is bulked up to 1.0 compared to its 0.66 value in C22.
For both planets, Bayesian evidence has been impacted: for HD 189733~b the Bayesian evidence decreased by $\approx$2.5 points (from 147.5 for C22 to 145.28 here) and increased by 10 points for HD 209458~b (from 198.76 for C22 to 218 here).
WASP-4 b, WASP-74 b, HAT-P-2 b and WASP-19 b were clearly subsolar under free chemistry assumptions in the retrieval calculation, but using here the disequilibrium FRECKLL model moved their metallicity toward solar (for WASP-4 b) to strongly supersolar (WASP-74 b, HAT-P-2 b and WASP-19 b), impacting as well their C/O ratio. 
For the most extreme example, WASP-74 b displayed a subsolar metallicity of log(Z)=-2.1 and a C/O ratio of 0.7 in free chemistry; here we have obtained a supersolar metallicity of log(Z)=1.74 and a C/O ratio of 0.47. 
Both chemistry assumptions result in very close Bayesian evidence values of 202.8 (free) and 203.29 (FRECKLL-only).
In the same vein, planets with an around solar metallicity in free chemistry (Kepler-13~A~b, TrES-3~b, WASP-43 b and WASP-77 A b) are, with FRECKLL, moved to supersolar (Kepler-13~A~b and TrES-3~b) and subsolar (WASP-43~b and WASP-77~A~b) metallicity conditions. 
While the C/O ratio of WASP-77~A~b, for instance, is roughly the same between the two methods (0.4 for free and 0.5 for disequilibrium), its metallicity has been changed from solar (in free) to subsolar (log(Z)$\sim$-1) using the disequilibrium model (Figure \ref{fig:emission_CO_Z}).
On the contrary, for TrES-3~b, free chemistry retrieval depicts this planet with a slightly subsolar metallicity (log(Z)$\sim$-0.4) and a C/O ratio of 0.2, but using FRECKLL as chemical model, this planet is here described with a supersolar metallicity ($\sim$0.8) and a higher C/O ratio of 0.5. 
As previously, taking into account disequilibrium chemistry in those four cases does not significantly impact the Bayesian evidence. Disequilibrium retrievals carried out here imply an increase of 1 point on the Bayesian evidence for Kepler-13 A b, an increase of 2 points for WASP-77 A b, and a decrease of 5 points for TrES-3~b, and 1 point for WASP-43 b compared to results from C22 free chemistry retrievals. 
Nonetheless, we note that the results of retrievals using FRECKLL for eclipse and transit data (Figure \ref{fig:emission_CO_Z} and Figure \ref{fig:transmission_CO_Z}, respectively) are consistent with each other, describing each planet in the same log(Z)-C/O region in both geometries, except for HD 209458~b where C/O ratio varies from $\sim$1 in eclipse to 0.2 in transit. 
Metallicity and C/O ratio are therefore model dependent, and this makes results interpretation in the context of planet formation difficult. 
\\

\begin{figure*}
   \centering
   \includegraphics[width=0.9\textwidth]{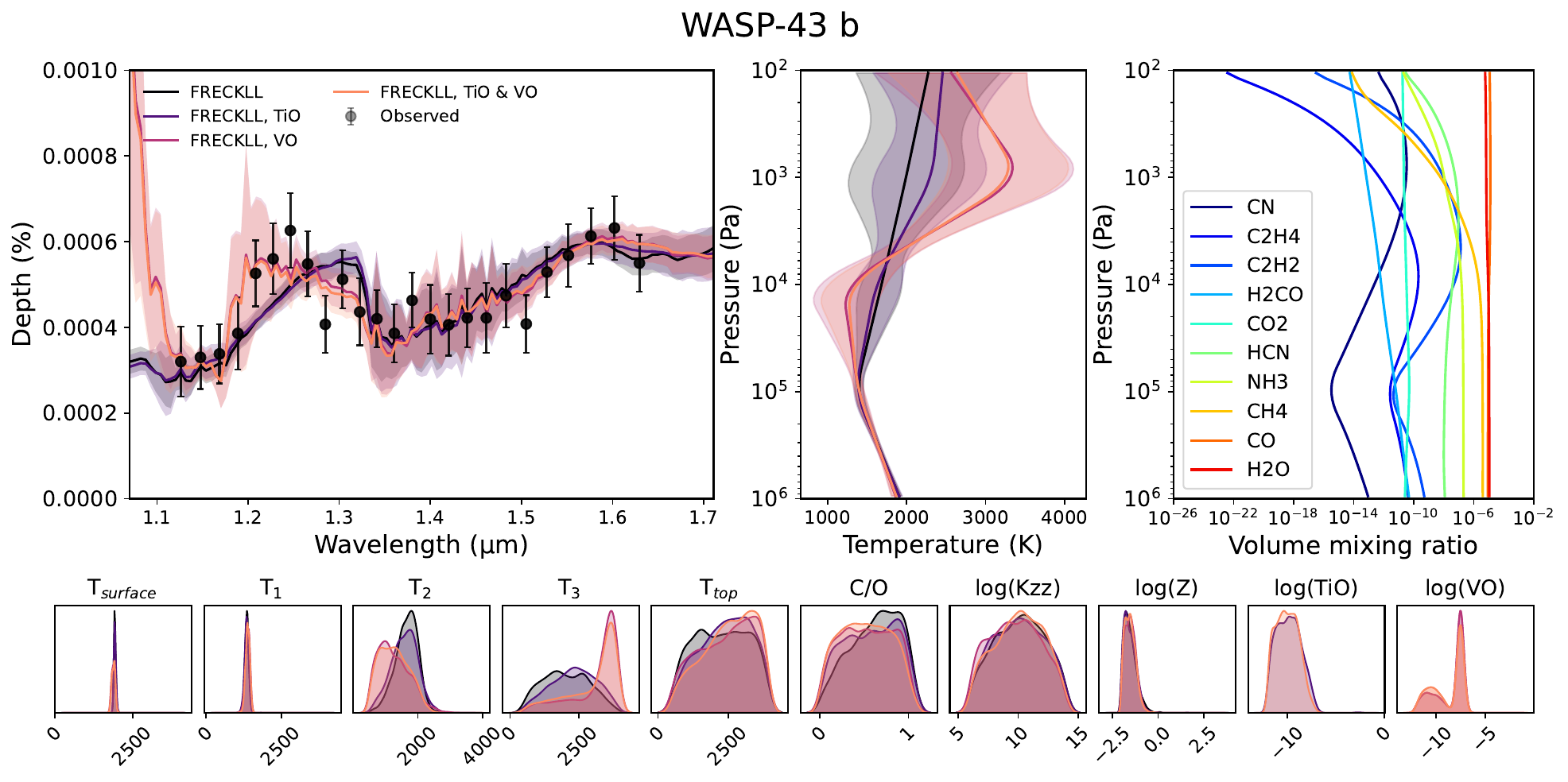}
   \caption{Detailed retrieval results for WASP-43 b: Upper row presents (from left to right) the fitting spectra for each retrieval configuration tested through this population study, the retrieved temperature profile, as well as the chemical structure of the planet (only for the ``FRECKLL-only'' retrieval.
   The lower row presents the posterior for each test, as normalised distribution. Equivalent figures for the rest of the population sample is available in Appendix \ref{app:indiv_eclipse} for eclipse retrieval results and Appendix \ref{app:indiv_transit} for transit retrieval results.}
              \label{fig:WASP43b}%
\end{figure*}

\begin{figure*}
   \centering
   \includegraphics[width=0.9\textwidth]{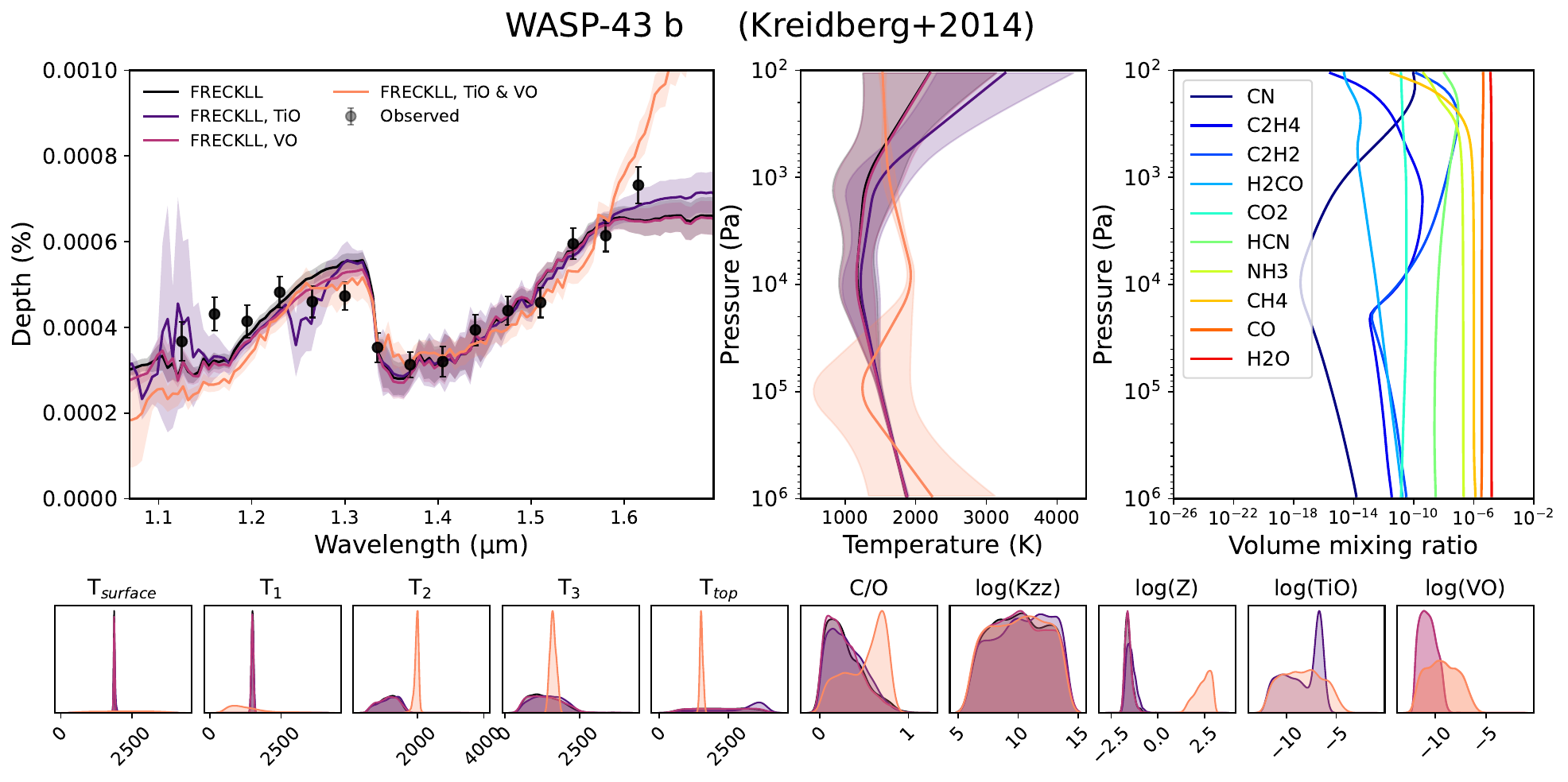}
   \caption{Detailed retrieval results for WASP-43 b using data reduced by \cite{Krei:14_WASP43b}: Same as Figure \ref{fig:WASP43b}. Here, retrieved parameters, prior setup and stellar and planetary configurations have been kept the same as before, we have only change the data for the one reduced by \cite{Krei:14_WASP43b}.}
              \label{fig:WASP43b_Kreidberg}%
\end{figure*}

\begin{figure*}
   \centering
   \includegraphics[width=0.9\textwidth]{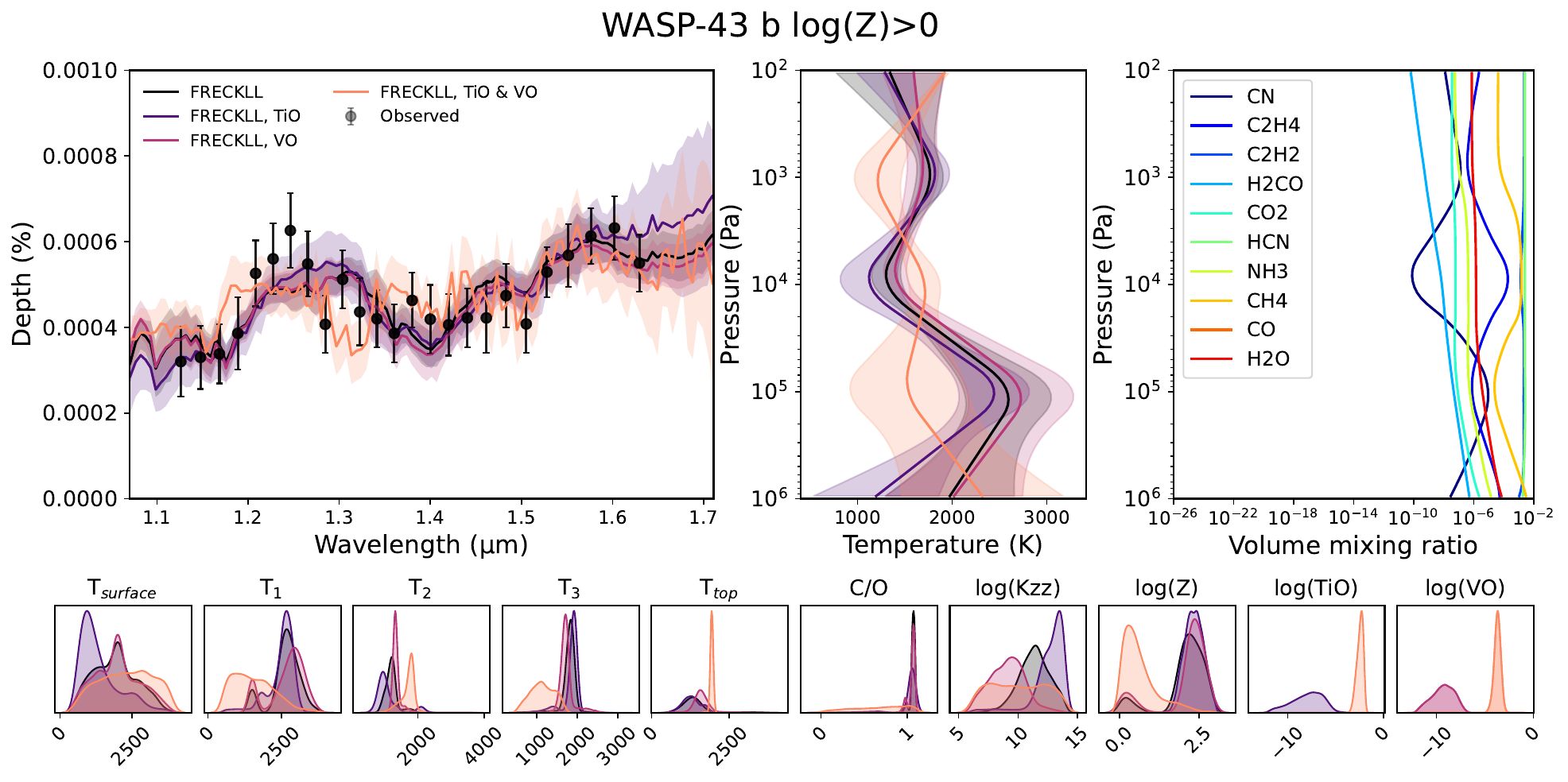}
   \caption{Detailed retrieval results for WASP-43 b with a restrictive metallicity (log(Z)) prior: Upper row presents (from left to right) the fitting spectra for each retrieval configuration tested through this population study, the retrieved temperature profile, as well as the chemical structure of the planet (only for the ``FRECKLL-only'' retrieval.
   The lower row presents the posterior for each test, as normalised distribution.}
              \label{fig:WASP43b_emission_logZpos}
\end{figure*}

\begin{figure*}
   \centering
   \includegraphics[width=0.9\textwidth]{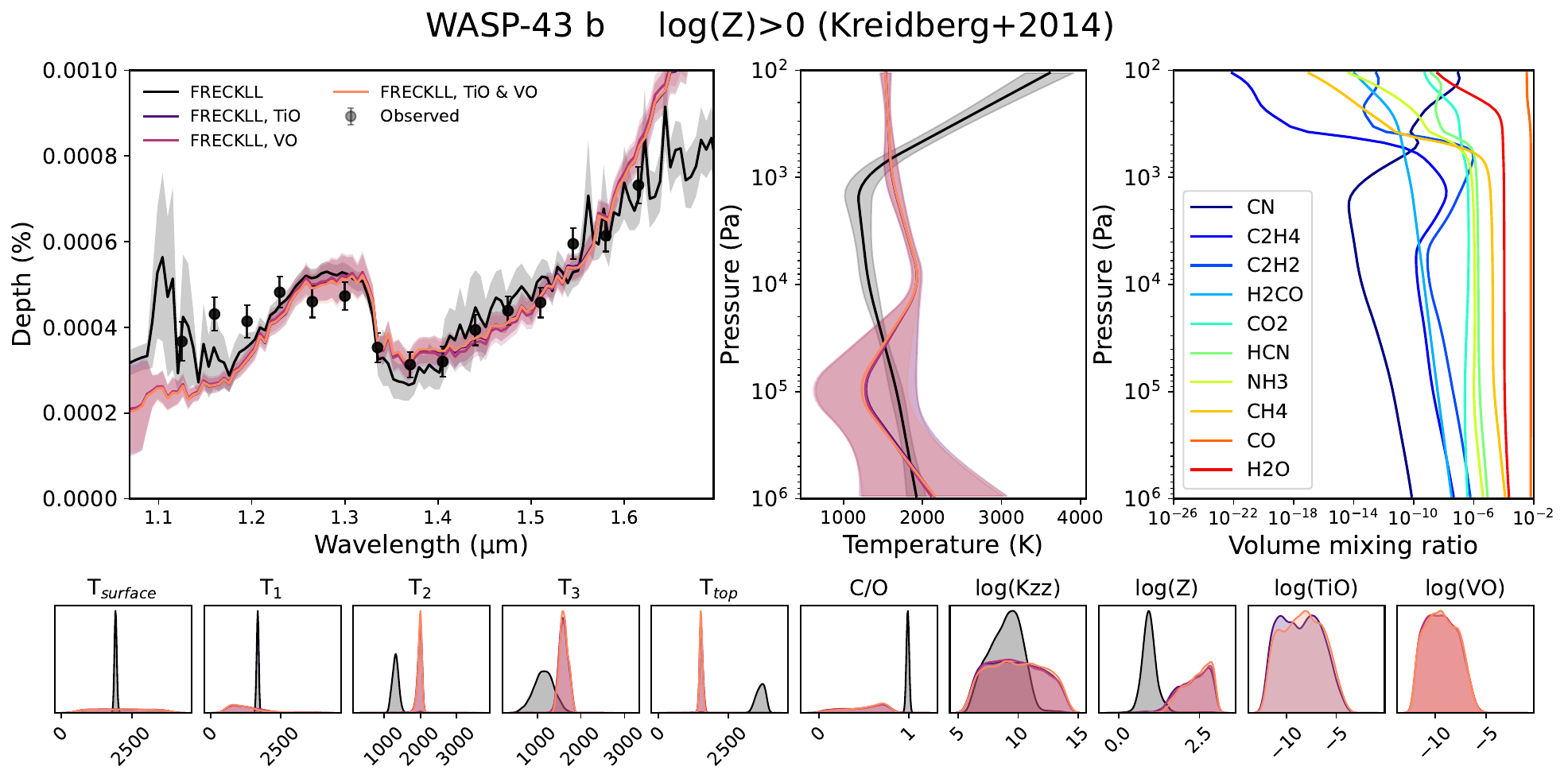}
   \caption{Detailed retrieval results for WASP-43 b using data reduced by \cite{Krei:14_WASP43b} with a restrictive metallicity (log(Z)) prior: Same as Figure \ref{fig:WASP43b_emission_logZpos}. Here, retrieved parameters, prior setup and stellar and planetary configurations have been kept the same as before, we have only change the data for the one reduced by \cite{Krei:14_WASP43b}.}
              \label{fig:WASP43b_Kreidberg_logZpos}%
\end{figure*}

The use of a disequilibrium chemical model during retrieval calculations highlights the fact that the conclusions drawn from HST observations are first dependent on the retrieval parameters, priors and more broadly on the model used, and then dependent on the reduction method.
This is shown through the example of the planet WASP-43~b (Figures \ref{fig:WASP43b}-\ref{fig:WASP43b_Kreidberg_logZpos}), using the reduced data from C22 and \cite{Krei:14_WASP43b} (hereinafter K14). 
Retrieved parameters, prior setup and stellar and planetary configurations (see Table \ref{tab:retrieval_settings}) have been kept the same across both dataset to ensure consistency during the comparison. 
In addition, we have encompassed the Bayesian evidence of each retrieval tests carried to address this point in Table \ref{tab:WASP43b_C22_K14}.

Overall, the thermal profiles for the two reduced dataset are consistent, especially for ``FRECKLL-only'', ``FRECKLL-TiO'' for both dataset, as well as ``FRECKLL-VO'' for K14 (Figures \ref{fig:WASP43b} and \ref{fig:WASP43b_Kreidberg}). 
The above-mentioned retrieval tests result in a thermal profile describing a decrease of temperature with altitude up to $\sim$10$^{5}$Pa for C22, and up to $\sim$10$^{4}$Pa for K14, in addition to a second thermal inversion at $\sim$10$^{3}$Pa for K14. 
For lower pressures, the temperature increases with altitude, suggesting two distinct atmospheric layers. 
This increase is within larger error bars, which suggests that greater care should be taken when reading these temperature and pressure profiles.  
However, ``FRECKLL-VO'' and ``FRECKLL-TiO-VO'' retrievals for C22 dataset depict a three-layer-atmosphere in the thermal profile: (bottom-up) a first layer of temperature decreasing up to $\sim$10$^{4}$Pa, a second layer of temperature increasing from $\sim$10$^{4}$Pa to $\sim$10$^{3}$Pa, and a third layer of temperature decreasing from $\sim$10$^{3}$Pa to the top of the modelled atmosphere (10$^{2}$Pa). 
Moreover, the retrieval test ``FRECKLL-TiO-VO'' for K14 dataset, presents a thermal profile entirely different (Figure \ref{fig:WASP43b_Kreidberg}). 
The thermal profile in the case of ``FRECKLL-TiO-VO'' for K14 depicts an inversion of temperature (from decrease to increase of temperature with the decreasing pressure) at 10$^{5}$Pa and a second one at 10$^{4}$Pa from which the temperature slightly decreases with altitude. 
For both, the temperatures at the bottom of the atmosphere are constrained, within a small error envelope  and roughly equivalent. 
Despite these above-named consistencies, the chemical structure for the reference ``FRECKLL-only'' run is largely impacted. 
In particular for the C$_2$H$_2$, C$_2$H$_4$, and CN profiles, for which the quenching level has increased by up to one decade of pressure.
As expected, the C/O ratio remains very challenging to be retrieved because HST observations with the WFC-3 instrument (whatever the reduction methods) lack sensitivity to C-bearing species due to a limited spectral range, this is particularly the case for the C22 dataset (Figure \ref{fig:WASP43b}), for which each retrieved tests depict a broad posterior distribution for the C/O ratio.
In the case of K14 dataset, this posterior distribution for the C/O ratio tends to a value of about 0 (i.e., a carbon-poor atmosphere), for most of the retrieval tests, except the ``FRECKLL-TiO-VO'' run presenting a C/O ratio tending to 1 (Figure \ref{fig:WASP43b_Kreidberg}). 
Reduced data from K14 and C22 present a large difference in resolution (data binning), as well as larger errobars for C22 than K14. 
Difference in data binning might be one of the sources for the difference in retrieval results between C22 and K14 datasets. 
However, population studies, such as \cite{Tsia:18} and \cite{Chan:22_5keyQuestions} found that, with HST data, difference in data binning does not impact much their retrieval results (when the low/high resolution reductions come from the same pipeline), so other effects could also contribute to the difference in retrieval results between C22 and K14 datasets.

Retrieved results from both reduced datasets agree on the metallicity (log(Z)$\sim$-2), considering WASP-43~b as a subsolar-metallicity planet.
However, those retrieval results, whether with C22 or K14 reduced datasets, are in strong opposition to the results obtained by those previously mentioned studies, e.g., 0.8$\times$solar in log for C22  and 0.18$\times$solar in log for K14 (converted from the value of 1.5$\times$solar in linear in K14 paper)). 
To allow a consistent comparison of the effect of the FRECKLL model during retrieval calculations, we have carried out additional retrieval tests on both reduced datasets for WASP-43~b, with the same set of retrieval runs, namely a ``FRECKLL-only'', a ``FRECKLL-TiO'', a ``FRECKLL-VO'', and a ``FRECKLL-TiO-VO'' run, but constraining the metallicity prior to the supersolar region, from 1 to 10$^{3}$ (to solar, in log).
Results from these additional retrievals are shown in Figure \ref{fig:WASP43b_emission_logZpos} and Figure \ref{fig:WASP43b_Kreidberg_logZpos}.

\begin{table*}
    \caption[]{Bayesian evidence for each retrievals carried on WASP-43~b eclipse observed spectra reduced by C22 and K14.}
    \label{tab:WASP43b_C22_K14} 
    \begin{tabular}{lcccccccc}
        \hline
                & FRECKLL & FRECKLL & FRECKLL & FRECKLL & FRECKLL & FRECKLL & FRECKLL & FRECKLL \\
  &   &    &   &    &  log(Z)>0 & log(Z)>0   & log(Z)>0   & log(Z)>0    \\  
  &   &  TiO & VO &  TiO+VO &   &   TiO &  VO & TiO+VO \\
                \hline
                \textbf{WASP-43~b, C22} & & & & & & & \\
        ln(E)    & 194.9 & 194.46 & 194.44 & 194.11 & 190.35 & 184.47 & 188.97 & 185.39 \\
                \hline    
                \textbf{WASP-43~b, K14} & & & & & & & \\
        ln(E)    & 111.23 & 111.65 & 110.38 & 99.80 & 105.21 & 100.98 & 100.35 & 100.19 \\
                \hline    
    \end{tabular}
\end{table*}    

Overall, for both reduced datasets, restricting the metallicity priors largely impacts the Bayesian evidences of the retrievals (see Table \ref{tab:WASP43b_C22_K14}), with a $\Delta$ln(E) up to 10 points lower when the prior is restricted to the positive region. 
With a metallicity prior constraint to the supersolar region for the C22 dataset, we obtain a bi-modal distribution of metallicity, with a most probable value of 2.5 to solar (in log) and a second probable value of 0 to solar (in log, i.e., equal to solar metallicity) for all retrieval tests, except for the ``FRECKLL-TiO-VO'' retrieval (for which the metallicity posterior distribution peaks at zero, suggesting an overly constraining priority in this case). 
Despite the consistency with the results obtained by C22 for WASP-43~b metallicity, this close proximity to one of the prior boundaries suggests that the model did not have enough freedom to explore suitable(s) value(s) to properly constrain the metallicity. 
Concerning the K14 dataset, metallicity distribution is constrained only in the case of ``FRECKLL-only''. 
For the rest of the retrieval tests, the retrieved metallicity displays a sparse distribution around 2.5 to solar (in log). 
In addition, ``FRECKLL-only'' retrieval results in a thermal profile approaching the vertical variation of the results obtained by the full-prior of metallicity retrieval for either C22 and K14 (Figures \ref{fig:WASP43b} and \ref{fig:WASP43b_Kreidberg}). 
Constraining the metallicity prior breaks the consistency in the thermal structure between the two data reduction methods, leads to thermal profiles with a lot of temperature inversions, one temperature inversion per decade of pressure (in particular for the C22 dataset), and this is even reinforced when considering TiO or/and VO in the chemical calculations. 
Those thermal profiles also impact the chemical structure of the planet for both reduced datasets, particularly in the case of C22.

Therefore, with WASP-43~b datasets produced by two different reduction methods, we highlight that the retrieval calculations are lightly impacted by the reduction method, but above all sensitive to the retrieval prior. 
With too much constraints on the parameter space, the model is not able to properly investigate and ``pulls'' on the prior by determining retrieved values on the prior boundary. 
Uncertainties remain concerning the actual metallicity of this planet as our results are opposed to the previous studies. 
Both K14 and C22 retrieval calculations of WASP-43~b retrieved the abundances of chemical species following a free model (e.g., constant-with-altitude profiles of abundance). 
Since these two retrieval calculations differ in the method used to reduce the observations, as well as the retrieval model used (although following the same calculation assumptions), obtaining supersolar metallicity for WASP-43~b for both K14 and C22 cases reinforces the conclusion that this planetary atmosphere probably possesses such metallicity, even if the value obtained varies from one method to another (0.18$\times$solar for K14 and 0.8$\times$solar for C22). 
However, using the disequilibrium chemical model FRECKLL, the planetary atmosphere of WASP-43~b appears as a subsolar metallicity atmosphere, with a well enough consistent thermal and chemical structures across both C22 and K14 reduced dataset. 
\\

While using a kinetic chemical model to calculate the chemical structure during retrieval calculation, contributions of TiO and VO (not implemented in FRECKLL, they are described as constant profiles) do not improve the retrieved results for almost all planets considered in this study. 
This statement is consistent with the design of our exoplanet sample (i.e., no inversion in the thermal profile, no without TiO and/or VO spectral signature, and an equilibrium temperature under 2500 K, as described in Section \ref{sec:methodo}).
Table \ref{tab:eclipse_retrieval_results} encompasses the retrieval results for each test and each planet studied in the present paper, showing that only for two planets, TrES-3~b and WASP-77~A~b, retrieval results benefit from the addition of TiO and VO contributions (by obtaining a $\Delta$ln(E)$>$5) compared with the ``FRECKLL-only'' retrieval results. 

In both cases, the ``FRECKLL-TiO'' retrieval is associated to the higher Evidence, with ln(E)$=$117.12 for TrES-3~b  ($\Delta$ln(E)$=$6 compared to ``FRECKLL-only'' case) and ln(E)$=$199.5 for WASP-77~A~b ($\Delta$ln(E)$\approx$5 compared to ``FRECKLL-only'' case). 
However, regarding retrieval results in the case of TrES-3~b (Figure \ref{fig:TrES3b_indiv}), ``FRECKLL-TiO'' configuration does not clearly improve the convergence of the retrieved parameters compared to other retrieval configurations.
All temperature points, C/O ratio, and metallicity posterior distributions present equivalent variations whatever the retrieval configurations considered, without any clear convergence toward a precise value. 
Metallicity posterior describes a slightly bi-modal distribution, with a preferential value of log(Z)=1 implying a supersolar atmosphere. 
C/O ratio could be any value between 0 and 1, confirming the challenge to constrain such planetary parameters with HST observations. 
Thermal structures are not constrained as well for all cases, with a large 1000-K-wide envelope of probability on both sides of the profiles. 
The resulting atmosphere consists of a bottom layer in which temperature is decreasing with altitude up to 10$^{5}$Pa, then a layer of a slow increase of temperature with altitude (except the ``FRECKLL-TiO-VO'' case, displaying a slow decrease of temperature with altitude, with a smaller gradient than the bottom layer), and a top layer with a large positive gradient of temperature.  
Moreover, we would like to draw the attention of the reader that two teams produced widely different datasets for TrES-3 b \citep{Chan:22_5keyQuestions, Ranj:2014}, likely due to the staring mode used for this observation. 
Therefore, despite the improvement in term of Evidence, we preferred to focus on the conservative model for this planet, i.e., the ``FRECKLL-only'' results.

On the contrary for WASP-77 A b--which is a more recent observation in scanning mode that allows us to discuss the presence of TiO more robustly--, taking into account TiO as a chemical contributor for the retrieval allows us to better constrain (i) the thermal structure, with narrow error bars around the temperature points, (ii) the metallicity, (iii) as well as the TiO mean abundance, but the C/O ratio remains a challenge for these particular retrieval results (Figure \ref{fig:WASP77Ab_indiv}). 
Among all retrieval configurations, ``FRECKLL-TiO'' is the only one resulting in a subsolar metallicity atmosphere with log(Z)$\approx$-1, whereas the three other configurations describe WASP-77~A~b as a supersolar metallicity atmosphere, all agreeing on log(Z)$\approx$2.5. 
This disagreement between ``FRECKLL-TiO'' and other configurations of retrieval is also reflected in the C/O ratio posterior distribution: all retrieval tests result in an equivalent posterior of C/O ratio, which would seem to converge toward 0.2.
Adding only TiO as a chemical contributor implies a large and imprecise distribution of C/O value, with a maximum value of 0.48 (that could not really be identified as a ``most probable'' value regarding the shape of the posterior). 
As in the TrES-3~b case, adding a degree of complexity by adding the chemical contribution of TiO makes the constraints on metallicity and C/O ratio even weaker, and does not really allow us to conclude on one value being more plausible than another. 
However, adding TiO permits to better constrain the bottom part of the retrieved thermal profile for the modelled atmosphere. 
Indeed, the three lowest temperature points (T$_{surface}$ to T$_2$) precisely agreed toward a temperature value, with only a 250-K-wide envelope of uncertainties, when other configurations to retrieve this planet predict a 700-K error on either side of the thermal profile. 
Overall, all retrieval configurations agree about the shape, gradient and temperature inflexion points for the top layer of the atmosphere (from 10$^{3}$ to 10$^{2}$Pa), but in the ``FRECKLL-TiO'' case, the topmost temperature point T$_{top}$ posterior follows a wide distribution, and is less constrained.  
\\

While using the FRECKLL kinetic chemical model during the retrieval calculations, retrieved temperature profiles from transit and eclipse observations are almost consistent with each other for the lowest layers of the modelled atmosphere. 
To verify the consistency between eclipse and transit spectra retrieval while considering the disequilibrium chemistry assumptions, we have carried out two kinds of tests for a reduced set of planets (HD 189733~b, HD 209458~b, WASP-43~b, and WASP-74~b).
Both retrieval tests only use FRECKLL, with either an isothermal profile for the retrieved temperature calculations, or a 5-point thermal profile (see Figures \ref{fig:HD189733b_transmission} to \ref{fig:WASP74b_transmission}).
The use of an isothermal or a 5-point thermal profile largely impacts the spread of the error of several parameters such as: the planet radius, the C/O ratio, as well as the metallicity, for which the width of the posterior distribution is largely impacted. 
Planet radii are, for all cases, consistent between isothermal and 5-point thermal profile retrieval, with the smallest error envelope, in the case of the isothermal profile retrieval. 
Transit observations are firstly sensitive to scale height, which is at first order controlled by the radius, the planet mass and afterwards controlled by the mean temperature. 
Using a 5-point profile provides more flexibility, but it is also more complex (i.e., it leads to larger uncertainties on the T-p profile). 
As a consequence, the planet radius is better constrained by the isothermal assumption. 
Concerning C/O ratio and metallicity, the use of an isothermal profile or a 5-point profile leads to diametrically opposite results: if the C/O ratio converges toward $\approx$1 for one kind of temperature profile, it converges toward $\approx$0 for the other, and the same result is obtained with metallicity (except for WASP-74~b) that varies from largely subsolar (log(Z)$<$-1) to largely supersolar (log(Z)$>$1).
In the case of WASP-74~b (Figure \ref{fig:WASP74b_transmission}), metallicity describes a bi-modal posterior distribution for the isothermal retrieval, with a maximum value at log(Z)=2.04 (see Table \ref{tab:transit_retrieval_results}), consistent with the fairly constrained value obtained using a 5-point thermal profile (log(Z)=2.21). 

Concerning the thermal profile, the use of 5 inflection points to retrieve the temperature profile allows for better consistency between eclipse and transit observation inversion for the lower part of the modelled atmosphere -- however, this is not systematic for all planets considered in our sample --, whereas the upper part of the atmosphere depicts an isothermal variation with altitude, consistent in value with the transit isothermal retrieval. 
In the case of HD 189733~b (Figure \ref{fig:HD189733b_transmission}), the overall shape as well as the value of the temperature profile of the transit 5-point retrieval is consistent with the eclipse retrieval results in the case of log(Z)$>$0 (purple curves on Figure \ref{fig:HD189733b_emission}). 
Within the error bars, we have obtained a bottom temperature between 2200~K and 2500~K considering both eclipse and transit observations, with a top layer atmosphere around 1000~K, but the inversion level is deeper in the atmosphere for the transit retrieval: at 10$^{4}$Pa, whereas the inversion occurs at 10$^{3}$Pa for eclipse observation. 
In addition to a consistency between eclipse and transit observation inversion for the thermal profile in the case of HD 189733~b, the eclipse log(Z)$>$0 retrievals and the transit 5-point retrieval are consistent with each other for the metallicity (ensured by the restrictive prior for this parameter), and for the C/O ratio (also shown by Figures \ref{fig:emission_CO_Z} and \ref{fig:transmission_CO_Z}). 
Similar conclusions are drawn in the case of WASP-74~b (Figures \ref{fig:WASP74b_emission} and \ref{fig:WASP74b_transmission}), except for the C/O ratio: transit 5-point retrieval predicts a C/O ratio that tends toward 1, while eclipse log(Z)$>$0 retrievals agree for a C/O ratio that tends toward 0.

For the two other planets HD 209458~b and WASP-43~b, adding some complexity in the temperature inversion with the 5-point retrieval only permits obtaining very close values for the very bottom temperature point of the modelled atmosphere (2000 to 2500~K). 
Higher in the modelled atmosphere, the 5-point retrieval thermal profile joins the variation and the temperature values of the temperature profile obtained with the isothermal retrieval, further away from the results predicted by the eclipse retrieval configurations. 
Contrary to HD 189733~b and WASP-74~b, transit isothermal retrieval implies consistent results with the eclipse retrieval results for metallicity (log(Z)$\approx$-2), while the transit 5-point retrieval induces supersolar metallicity atmosphere and breaks consistency analysis between eclipse and transit for a particular planet.

\section{Discussion and Conclusions}
\label{sec:discussion}

In this paper, we have tested the TauREx-FRECKLL coupled model on Hubble Space Telescope WFC3 observations of 10 hot-Jupiter atmospheres, in the attempts to better constrain the metallicity, C/O ratio, thermal structure and, above all, the chemistry structure of those planets by using a more realistic chemistry hypothesis for spectral observations inversion. 
The TauREx-FRECKLL coupled model allows to take into account disequilibrium kinetic chemistry calculations -- thanks to the plugin to FRECKLL -- on fly of the state-of-the-art retrieval calculations made by TauREx-3.1 for exoplanets observations. 

To this purpose, we have conducted a series of retrieval experiments, considering the entire population, as well as specific planets to study (i) the impact of the disequilibrium chemistry hypothesis on the metallicity and C/O ratio inferring, (ii) the impact of the reduction method of HST/WFC3 observations, (iii) the impact of taking into account refractory elements TiO and VO under the disequilibrium chemistry hypothesis, and finally (iv) to check if the disequilibrium chemistry hypothesis can lead to a reconciliation between transit and eclipse retrieval results for a specific planet. 

(i) Considering the whole exoplanet population selected for this study, the disequilibrium chemistry hypothesis can strongly impact the conclusion drawn from the analysis of HST data under other and simpler chemistry hypotheses (as free or equilibrium, for example). 
Metallicity and C/O ratio values retrieved from transit and eclipse, while consistent with each other, are in some cases in opposition with previous studies using the same data. 
Thus, the characteristics retrieved with the disequilibrium hypothesis can have consequences for planet formation studies, by changing the group to which the considered exoplanet belongs.

(ii) Then, we have explored the impact of the reduction method on the retrieval of HST/WFC3 observations of WASP-43~b with our TauREx-FRECKLL coupled model. 
We have used reduced data from \cite{Krei:14_WASP43b} (K14) and \cite{Chan:22_5keyQuestions} (C22). 
Although both datasets agreed on the metallicity (not in value but both studies depicted the atmosphere of WASP-43~b as supersolar), our disequilibrium chemistry retrievals result in a subsolar metallicity for both datasets.
Disequilibrium retrievals conducted in the present paper are consistent with each other, for both datasets, in metallicity value, thermal structure and slightly in chemistry structure considering the simpler retrieval configurations (i.e., FRECKLL-only assumption).  
However, constraining retrieval prior of metallicity to try to match previous conclusions from both K14 and C22 implies an inconsistency in terms of metallicity distribution between those two datasets, and the retrieved metallicity distribution is too close to the limit values of the parameter space.  
Hence, using a disequilibrium chemical model during retrieval calculations highlights the fact that the conclusions drawn from HST observations are first dependent on the retrieval parameters, priors and more broadly on the model used, and in a lower extent dependent on the reduction method.

(iii) Considering refractory elements TiO and VO do not improve the goodness of the fit for most of the exoplanets studied in this study when using a kinetic chemical model to calculate the chemical structure during retrieval calculation. 
This is expected since most planets belong to the hot Jupiter category (i.e., below the metal-driven thermal inversion regime, C22).
Despite that TiO and VO have been previously detected in eclipse spectra using HST data and have been shown in the literature to help constrain the recovered metallicity and C/O ratio, and thus better infer planet formation scenario, we noticed a sensitive improvement of the fit only for two exoplanets TrES-3~b and WASP-77~A~b (i.e., an increase of minimum 5 points for ln(E)), and only with the addition of TiO. 
While taking into account TiO doesn't impact the convergence of retrieved parameters for TrEs-3~b, the addition of TiO actually impacts the convergence of both WASP-77~A~b thermal and the chemical structures. 
However, in this case, C/O ratio and metallicity are in opposition compared to the results of other retrieval experimentation.

(iv) In transit view, spectral observations are primarily sensitive to the atmospheric scale height, depending on the planet radius, planet mass, and the mean temperature. 
In the present study, by comparing eclipse and transit data retrievals, we have obtained retrieved temperature profiles consistent between both views (i.e., eclipse and transit) for the lowest layers of the modelled atmosphere.
This consistency has been obtained in the case of 5 temperature-pressure points discretization to describe the thermal profile, allowing up to 3 inflection points for the thermal structure of the planet (i.e., tests using an isothermal profile for the retrieval still depicted a bias of about a thousand kelvin compared to retrieval using eclipse data). 
Hence, up to 10$^{4}$Pa, retrieved temperatures from transit data are within the error bars of those obtained from eclipse data. 
Adding complexity to the thermal profile while retrieving transit observations is also necessary for JWST observations: \cite{Schl:24} found that a two-point profile is sufficient to retrieve the known atmospheric parameters, while
under the presence of an atmospheric temperature inversion, it is necessary to use an even more complex temperature profile. 
For the highest part of the modelled atmospheres, temperature profiles follow an isothermal variation with altitude,  approaching the results obtained with the isothermal profiles and thus keeping the bias of nearly 1000 K compared to the results of analysis of eclipse observations. 
This latter result is consistent with the fact that eclipse and transit views do not probe the same region of the atmosphere.

Results from the use of a disequilibrium chemistry model coupled with the retrieval algorithm are highly dependent on the spectral resolution of the data. 
With this first application to real data, we demonstrate the feasibility of such complex retrievals, although the low spectral coverage of HST does not offer the possibility of obtaining stronger constraints than in previous studies assuming simpler chemistry assumptions (free or thermodynamic equilibrium). 
In a following study, it will be interesting to conduct an equivalent population study with James Webb Space Telescope observations, that allows a wider spectral coverage -- allowing to obtain more and different molecular signatures to constrain the thermal and chemical structures and thus the metallicity and C/O ratio of the atmosphere --  as well as a higher spectral resolution.

\begin{acknowledgements}
      Bardet and Venot acknowledge funding from Agence Nationale de la Recherche (ANR), project ``EXACT'' (ANR-21-CE49-0008-01). In addition, Venot acknowledges funding from the Centre National d’Études Spatiales (CNES).
      
      The authors acknowledge the exceptional computing support from Grand Equipement National de Calcul Intensif (GENCI) and Centre Informatique National de l'Enseignement Supérieur (CINES). 
      All the simulations presented here were carried out on the ADASTRA cluster hosted at CINES. 
      This work was granted access to the High-Performance Computing (HPC) resources of CINES under the allocations A0140110391, and A0160110391 made by GENCI. 
      The total number of computing hours used to carry out this study amounts to 874,207~hCPU. 
      All retrieval results and parameter files of each retrieval runs are available on the EXACT project website at \url{https://www.anr-exact.cnrs.fr/fr/retrieval/}. 
      All code used for generation of figures is available on GitHub at \url{https://github.com/debbardet/EXACT_plotter}.
\end{acknowledgements}

   \bibliographystyle{aa} 
   \bibliography{newfred}

\begin{appendix} 
\section{Sensitivity exploration on the dependence of the retrieved thermal profiles on the number of temperature points}
\label{app:sensitivity_npoints_Tp}

Setting the number of parameters necessary to retrieve and analyse spectral observations of exoplanets can be really difficult, in particular for the temperature vertical profile. 
\cite{Chan:21_phasecurve_WASP43b} have studied in detail the impact of retrieval parameters on the resulting temperature profile of WASP-43~b using HST eclipse data, especially the number of N-points to describe the temperature profile (see their appendix D.). 
In their study, they have found it very difficult to determine the ideal number of N-points. 
In \cite{Schl:24} study, the authors have shown that an isothermal temperature profile for analysing HST data is sufficient, but for JWST data, it is necessary to use at least 2 points. 
In the latter case, they have shown that adding more points does not necessarily bias the result.

\begin{figure*}
   \centering
   \includegraphics[width=0.9\textwidth]{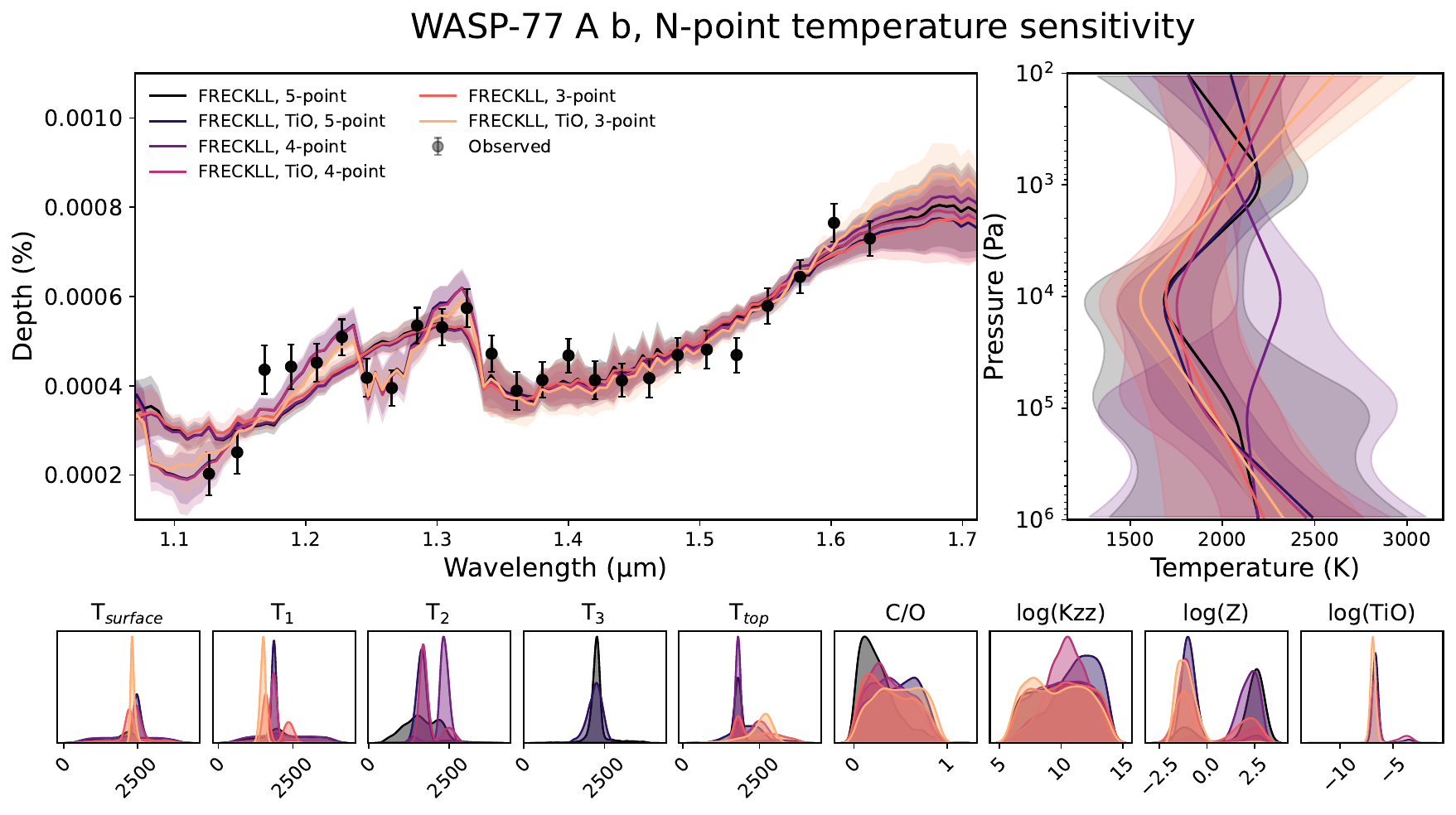}
   \caption{Detailed retrieval results for the sensitivity exploration on the dependence of the retrieval thermal profiles on the number of temperature points with the example of WASP-77~A~b with and without the addition of TiO. 
   Upper row presents (from left to right) the fitting spectra for each retrieval configuration tested through this population study, and the retrieved temperature profile.
   The lower row presents the posterior for each test, as normalised distribution.}
    \label{fig:Sensitivity_Npoint_temperature_profiles}
\end{figure*}

In the present population study of 10 hot-Jupiter exo-atmospheres, we have adopted a consistent retrieval methodology including, among others, a 5-point thermal profile to retrieve the temperature vertical structure.  
To explore the dependence of the resulting temperature profile on the number of N-points set as retrieval parameters, we have conducted additional retrieval calculations considering only 3 and 4 temperature points, with and without the addition of TiO as a constant abundance profile in the case of WASP-77 A b, and all the other retrieval parameters remain the same as described in Section \ref{sec:methodo}.
For the 3-point method, we have a unique inflection point at the fixed pressure of 10$^4$ Pa. 
For the 4-point method, we have set an additional inflection temperature point at the fixed pressure of 10$^5$ Pa. 
Results of those additional retrieval calculations are shown in Figure \ref{fig:Sensitivity_Npoint_temperature_profiles} and Table \ref{tab:sensitivity_Npoint_temperature_profile}. 

Whatever the number of points chosen for the temperature profile, adding the TiO compound to the retrieval parameters significantly improves the representation of the spectrum (Figure \ref{fig:Sensitivity_Npoint_temperature_profiles}), induces consistent temperature profiles between all cases (except for the ``FRECKLL, 4-point'' run), and all retrieval tests including TiO determine closed values for its abundance ($\approx$10$^{-7}$ in volume mixing ratio). 
In general, determination of the C/O ratio, as well as the metallicity, remains difficult, but some of those retrieval calculations depict consistent values. 
For instance, the C/O ratio of WASP-77 A b is estimated between 0.22 and 0.48, with this latter being determined with 3-point and 5-point temperature profiles, both including TiO. 
In addition, both ``4-point'' retrieval runs (i.e., with and without TiO), as well as the ``3-point'' retrieval run result in close C/O ratio values, evaluating it between 0.34 and 0.39. 
For the metallicity, it is more complex to draw any trend; depending on the case, WASP-77 A b is determined as solar to super solar in metallicity. 
Finally, for runs without TiO, the addition of points in the temperature profile impacts the Bayesian evidence by a reduction of only 1.3, while the reduction is 2.5 when TiO is taken into account. 

As \cite{Chan:21_phasecurve_WASP43b} and \cite{Schl:24}, we found that it is complex to really determine the optimum number of points to describe the retrieved temperature profile using the N-point methods, and in our case, with the example of WASP-77 A b, the addition of more points does not bias the results, but allows more freedom to the Multinest sampling and offers a more detailed description of the vertical thermal structure. 
This sensibility study will be interesting to conduct on JWST observations. 

\begin{table}
\caption{\label{tab:sensitivity_Npoint_temperature_profile} Retrieval results for the dependence of the retrieved thermal profiles on the number of temperature points}
\centering
\begin{tabular}{lcc}
Parameters & FRECKLL & FRECKLL \\
           &  & TiO \\
\hline
\hline

\textbf{5-point thermal profile} & & \\
log(TiO) & ... & -6.65$^{+0.37}_{-0.26}$ \\
C/O      & 0.22$^{+0.32}_{-0.17}$ & 0.48$^{+0.29}_{-0.35}$ \\
log(Z)   & 0.16$^{+0.90}_{-1.61}$ & -0.58$^{+1.28}_{-1.14}$ \\
ln(E)    & 195.31 & 199.5 \\ 
\hline

\textbf{4-point thermal profile} & & \\
log(TiO) & ... & -6.68$^{+2.74}_{-0.32}$ \\
C/O      & 0.39$^{+0.32}_{-0.29}$ & 0.34$^{+0.34}_{-0.24}$ \\
log(Z)   & 2.23$^{+0.56}_{0.91}$ & -1.03$^{+3.43}_{-0.38}$ \\
ln(E)    & 195.99 & 199.96 \\ 
\hline

\textbf{3-point thermal profile} & &  \\
log(TiO) & ... & -6.91$^{+0.32}_{-0.28}$ \\
C/O      & 0.36$^{+0.27}_{-0.38}$ & 0.48$^{+0.33}_{-0.34}$ \\
log(Z)   & -0.89$^{+3.44}_{-0.61}$ & -1.16$^{+0.69}_{-0.51}$ \\
ln(E)    & 196.64 & 202.05 \\ 
\hline
\end{tabular}
\end{table}

\section{Stellar spectra data}
\label{app:stellar_flux}

To obtain stellar spectra for each of the stars considered here, we use the referenced data from type F, G, K stars -- from the 
Virtual Planetary Laboratory, Spectral Database \& Tools database \url{http://depts.washington.edu/naivpl/content/spectral-databases-and-tools} -- closest to our sample stars.
When available, we also use data for the 1-120 nm part of the UV stellar spectra from the X-Exoplanets database (\url{http://sdc.cab.inta-csic.es/xexoplanets/jsp/homepage.jsp}) and the MUSCLES Treasury Survey database (\url{https://cos.colorado.edu/~kevinf/muscles.html}).
For instance, as HD 209458 is a G0 star \citep{Salz:15}, TrES-3 is a G4 star \citep{ODon:07}, and WASP-4, WASP-19 and WASP-77~A are G8 stars \citep[][respectively]{Tria:10, Hebb:10, Salz:15}, we use for those 5 stars the UV spectral irradiance of the Sun \citep{Thui:04} scaled to correspond to the radius and effective temperature of each of them. 

For missing spectral regions in the recorded data, we use a theoretical Kurucz spectrum model to model our star ensemble (\url{http://kurucz.harvard.edu/grids.html}). 
By selecting the grid file corresponding to a star with the closest metallicity [Fe/H] of each of our individual stars, we model a theoretical stellar flux as a function of the effective temperature and gravity (in log scale) of the desired star.

Then, we carry out a step of normalization of the theoretical and observed spectra to obtain the stellar data for each star used in the retrieval calculations, taking into account photodissociation.

\section{Individual planet analysis in eclipse}
\label{app:indiv_eclipse}

\longtab[2]{
\begin{landscape}
\begin{longtable}{l*{12}{p{0.079\textwidth}}}
\caption{Retrieval Results for the 10 Planets using Eclipse Spectra} 
\label{tab:eclipse_retrieval_results} \\
\hline 
Parameters & FRECKLL & FRECKLL & FRECKLL & FRECKLL & FRECKLL & FRECKLL & FRECKLL & FRECKLL & FRECKLL & FRECKLL & FRECKLL &  FRECKLL\\
  &   &    &   &    &  log(Z)>0 & log(Z)>0   & log(Z)>0   & log(Z)>0  & log(Z)<0 & log(Z)<0  & log(Z)<0   & log(Z)<0  \\  
  &   &  TiO & VO &  TiO+VO &   &   TiO &  VO & TiO+VO & &  TiO &  VO & TiO+VO \\
\hline
\hline
\endfirsthead
\caption{continued.} \\
\hline
Parameters & FRECKLL & FRECKLL & FRECKLL & FRECKLL & FRECKLL & FRECKLL & FRECKLL & FRECKLL & FRECKLL & FRECKLL & FRECKLL &  FRECKLL\\
  &   &    &   &    &  log(Z)>0 & log(Z)>0   & log(Z)>0   & log(Z)>0  & log(Z)<0 & log(Z)<0  & log(Z)<0   & log(Z)<0  \\  
  &   &  TiO & VO &  TiO+VO &   &   TiO &  VO & TiO+VO & &  TiO &  VO & TiO+VO \\
\hline
\endhead
\hline
\endfoot
\textbf{HAT-P-2~b} & & & & & & & & & & & & \\
log(TiO) & ... & ... & ... & ... & ... & -8.39$^{+2.96}_{-2.60}$ & ... & -8.14$^{+2.72}_{-2.79}$ & ... & -9.36$^{+2.24}_{-1.96}$ & ... & -9.30$^{+2.06}_{-2.12}$ \\
log(VO)  & ... & ... & ... & ... & ... & ... & -6.39$^{+1.91}_{-3.68}$ & -6.49$^{+1.72}_{-4.10}$ & ... & ... & -7.48$^{+0.43}_{-1.33}$ & -7.48$^{+0.44}_{-0.93}$ \\
C/O      & 0.52$^{+0.30}_{-0.31}$ & ... & ... & ... & 0.49$^{+0.30}_{-0.29}$ & 0.48$^{+0.31}_{-0.30}$ & 0.48$^{+0.31}_{-0.28}$ & 0.50$^{+0.30}_{-0.29}$ & 0.57$^{+0.31}_{-0.34}$ & 0.54$^{+0.34}_{-0.32}$ & 0.53$^{+0.35}_{-0.33}$ & 0.59$^{+0.31}_{-0.35}$ \\
log(Z)   & 1.73$^{+0.95}_{-3.10}$ & ... & ... & ... & 2.00$^{+0.75}_{-1.33}$ & 2.05$^{+0.71}_{-1.32}$ & 1.42$^{+1.19}_{-1.01}$ & 1.92$^{+0.79}_{-0.92}$ & -1.22$^{+0.72}_{-0.56}$ & -1.26$^{+0.81}_{-0.54}$ & -1.35$^{+0.64}_{-0.45}$ & -1.30$^{+0.69}_{-0.48}$  \\
ln(E)    & 218 & ... & ... & ... & 218.5 & 218.18 & 218.7 & 218.4 & 217 & 216.85 & 218.76 & 218.34 \\
\hline
\textbf{HD 189733~b} & & & & & & & & & & & & \\
log(TiO) & ... & ... & ... & ... & ... & -6.66$^{+3.68}_{-4.04}$ & ... & -7.27$^{+3.73}_{-3.40}$ & ... & -7.43$^{+3.19}_{-3.58}$ & ... & -7.08$^{+2.98}_{-3.63}$ \\
log(VO)  & ... & ... & ... & ... & ... & ... & -7.00$^{+3.12}_{-3.87}$ & -6.91$^{+2.97}_{-4.00}$ & ... & ... & -9.33$^{+2.77}_{-2.03}$ & -8.78$^{+2.33}_{-2.40}$ \\
C/O      & 0.54$^{+0.42}_{-0.43}$ & ... & ... & ... & 0.33$^{+0.39}_{-0.25}$ & 0.33$^{+0.36}_{-0.25}$ & 0.27$^{+0.39}_{-0.21}$ & 0.37$^{+0.41}_{-0.29}$ & 0.68$^{+030}_{-0.41}$ & 0.65$^{+0.33}_{-0.40}$ & 0.63$^{+0.33}_{-0.43}$ & 0.62$^{+0.33}_{-0.41}$ \\
log(Z)   & -1.05$^{+3.73}_{-0.84}$ & ... & ... & ... & 2.30$^{+0.56}_{-1.14}$ & 2.39$^{+0.48}_{-0.97}$ & 2.48$^{+0.43}_{-0.84}$ & 2.33$^{+0.54}_{-1.16}$ & -1.67$^{+0.59}_{-0.26}$ & -1.49$^{+0.82}_{-0.40}$ & -1.63$^{+0.66}_{-0.29}$ & -1.40$^{+0.80}_{-0.46}$  \\
ln(E)    & 114.70 & ... & ... & ... & 144.64 & 144.77 & 144.92 & 144.94 & 145.28 & 145.45 & 145.06 & 144.91 \\
\hline
\textbf{HD 209458~b} & & & & & & & & & & & & \\
log(TiO) & ... & -10.55$^{+1.19}_{-1.08}$ & ... & -2.64$^{+0.38}_{-0.53}$ & ... & ... & ... & ... & ... & ... & ... & ... \\
log(VO)  & ... & ... & -8.79$^{+0.23}_{-0.80}$ & -5.75$^{+0.42}_{-3.33}$ & ... & ... & ... & ... & ... & ... & ... & ... \\
C/O      & 1.01$^{+0.07}_{-0.04}$  & 1.03$^{+0.05}_{-0.06}$  & 0.98$^{+0.02}_{-0.02}$  & 0.60$^{+0.31}_{-0.41}$  & ... & ... & ... & ... & ... & ... & ... & ... \\
log(Z)   & -1.86$^{+0.24}_{-0.11}$  & -1.85$^{+0.25}_{-0.12}$  & -1.92$^{+0.14}_{-0.07}$  & -1.12$^{+0.63}_{-0.55}$  & ... & ... & ... & ... & ... & ... & ... & ... \\
ln(E)    & 218.42 & 216.16 & 219.26 & 220.31 & ... & ... & ... & ... & ... & ... & ... & ... \\
\hline
\textbf{Kepler-13~A~b} & & & & & & & & & & & & \\
log(TiO) & ... & -8.62$^{+2.50}_{-2.45}$ & ... & -8.23$^{+2.64}_{-2.87}$ & ... & ... & ... & ... & ... & ... & ... & ... \\
log(VO)  & ... & ... & -8.86$^{+2.26}_{-2.14}$ & -8.79$^{+2.39}_{-2.24}$ & ... & ... & ... & ... & ... & ... & ... & ... \\
C/O      & 0.51$^{+0.29}_{-0.31}$ & 0.50$^{+0.31}_{-0.29}$ & 0.47$^{+0.32}_{-0.28}$ & 0.50$^{+0.33}_{-0.29}$ & ... & ... & ... & ... & ... & ... & ... & ... \\
log(Z)   & 0.60$^{+1.42}_{-1.12}$ & 0.41$^{+1.42}_{-1.12}$ & 1.07$^{+1.32}_{-1.71}$ & 1.46$^{+1.03}_{-1.89}$ & ... & ... & ... & ... & ... & ... & ... & ... \\ 
ln(E)    & 115.47 & 115.49 & 114.92 & 115.10 & ... & ... & ... & ... & ... & ... & ... & ... \\
\hline
\textbf{TrES-3~b} & & & & & & & & & & & & \\
log(TiO) & ... & -7.06$^{+2.09}_{-3.29}$ & ... & -6.36$^{+2.03}_{-3.50}$ & ... & ... & ... & ... & ... & ... & ... & ... \\
log(VO)  & ... & ... & -9.02$^{+2.38}_{-2.10}$ & -8.87$^{+2.49}_{-2.22}$ & ... & ... & ... & ... & ... & ... & ... & ... \\
C/O      & 0.45$^{+0.35}_{-0.32}$ & 0.37$^{+0.36}_{-0.28}$ & 0.47$^{+0.33}_{-0.32}$ & 0.43$^{+0.33}_{-0.29}$ & ... & ... & ... & ... & ... & ... & ... & ... \\
log(Z)   & 0.63$^{+1.15}_{-2.07}$ & 0.60$^{+0.91}_{-1.19}$ & 0.60$^{+1.21}_{-1.72}$ & 0.96$^{+1.50}_{-1.05}$ & ... & ... & ... & ... & ... & ... & ... & ... \\
ln(E)    & 111.21 & 117.12 & 113.23 & 116.42 &  ... & ... & ... & ... & ... & ... & ... & ... \\ 
\hline 
\textbf{WASP-19~b} & & & & & & & & & & & & \\
log(TiO) & ... & ... & ... & ... & ... & -7.42$^{+3.02}_{-3.30}$ & ... & -7.73$^{+3.03}_{-3.11}$ & ... & -8.71$^{+2.46}_{-2.35}$ & ... & -8.05$^{+2.44}_{-2.71}$ \\
log(VO)  & ... & ... & ... & ... & ... & ... & -7.98$^{+2.67}_{-3.04}$ & -8.10$^{+2.86}_{-2.95}$ & ... & ... & -9.40$^{+2.01}_{-1.84}$ & -9.48$^{+2.06}_{-1.82}$ \\
C/O      & 0.52$^{+0.32}_{-0.35}$ & ... & ... & ... & 0.35$^{+0.39}_{-0.26}$ & 0.32$^{+0.37}_{-0.23}$ & 0.41$^{+0.35}_{-0.31}$ & 0.38$^{+031}_{-0.28}$ & 0.47$^{+0.36}_{-0.33}$ & 0.48$^{+0.36}_{-0.35}$ & 0.45$^{+0.34}_{-0.32}$ & 0.42$^{+0.35}_{-0.29}$ \\
log(Z)   & 1.62$^{+1.08}_{-2.96}$ & ... & ... & ... & 2.14$^{+0.63}_{-1.11}$ & 2.04$^{+0.74}_{-1.26}$ & 2.09$^{+0.64}_{-1.16}$ & 2.15$^{+0.65}_{-0.82}$ & -1.16$^{+0.74}_{-0.62}$ & -1.06$^{+0.67}_{-0.61}$ & -0.91$^{+0.61}_{-0.79}$ & -0.98$^{+1.33}_{-0.59}$ \\
ln(E)    & 172.16 & ... & ... & ... & 172.32 & 172.25 & 172.29 & 172.14 & 172.20 & 172.30 & 171.82 & 171.78 \\
\hline
\textbf{WASP-4~b} & & & & & & & & & & & & \\
log(TiO) & ... & -8.97$^{+3.23}_{-2.30}$ & ... & -6.40$^{+3.32}_{-4.79}$ & ... & ... & ... & ... & ... & ... & ... & ... \\
log(VO)  & ... & ... & -9.16$^{+5.73}_{-2.11}$ & -5.87$^{+3.45}_{-4.79}$ & ... & ... & ... & ... & ... & ... & ... & ... \\
C/O      & 0.47$^{+0.42}_{-0.35}$ & 0.43$^{+0.31}_{-0.31}$ & 0.58$^{+0.40}_{-0.41}$ & 0.55$^{+0.37}_{-0.40}$ & ... & ... & ... & ... & ... & ... & ... & ... \\
log(Z)   & -0.22$^{+1.47}_{-0.86}$ & 0.28$^{+1.25}_{-1.27}$ & -0.46$^{+1.97}_{-0.76}$ & 1.12$^{+1.11}_{-2.00}$ & ... & ... & ... & ... & ... & ... & ... & ... \\
ln(E)    & 121.20 & 120.60 & 120.36 & 120.15 & ... & ... & ... & ... & ... & ... & ... & ... \\ 
\hline
\textbf{WASP-43~b} & & & & & & & & & & & & \\
log(TiO) & ... & -9.89$^{+1.73}_{-1.60}$ &  ...  & -10.04$^{+1.57}_{-1.53}$ & ... & ... & ... & ... & ... & ... & ... & ... \\
log(VO)  & ... &  ...  & -7.65$^{+0.32}_{-3.49}$ & -7.75$^{+0.40}_{-3.46}$& ... & ... & ... & ... & ... & ... & ... & ... \\
C/O      & 0.65$^{+0.32}_{-0.43}$ & 0.59$^{+0.37}_{-0.43}$ & 0.53$^{+0.40}_{-0.40}$ & 0.52$^{+0.38}_{-0.38}$ & ... & ... & ... & ... & ... & ... & ... & ... \\
log(Z)   & -1.58$^{+0.51}_{-0.32}$ & -1.58$^{+0.48}_{-0.33}$ & -1.52$^{+0.41}_{-0.36}$ & -1.55$^{+0.41}_{-0.33}$ & ... & ... & ... & ... & ... & ... & ... & ... \\
ln(E)    & 194.9 & 194.46 & 195.44 & 194.11 & ... & ... & ... & ... & ... & ... & ... & ... \\ 
\hline
\textbf{WASP-74~b} & & & & & & & & & & & & \\
log(TiO) & ... & ... & ... & ... & ... & -7.03$^{+3.05}_{-3.55}$ & ... & -7.02$^{+3.21}_{-3.81}$ & ... & -8.88$^{+2.66}_{-2.25}$ & ... & -5.60$^{+3.43}_{-4.74}$ \\
log(VO)  & ... & ... & ... & ... & ... & ... & -7.55$^{+2.94}_{-3.20}$ & -7.28$^{+2.90}_{-3.31}$ & ... & ... & -9.56$^{+6.61}_{-1.81}$ & -8.39$^{+4.45}_{-2.67}$ \\
C/O      & 0.47$^{+0.45}_{-0.35}$ & ... & ... & ... & 0.40$^{+0.33}_{-0.29}$ & 0.36$^{+0.34}_{-0.27}$ & 0.34$^{+0.34}_{-0.26}$ & 0.34$^{+0.35}_{-0.25}$ & 0.76$^{+0.22}_{-0.50}$ & 0.69$^{+0.28}_{-0.48}$ & 0.72$^{+0.25}_{-0.46}$ & 0.62$^{+0.32}_{-0.43}$ \\
log(Z)   & 1.74$^{+0.98}_{-3.29}$ & ... & ... & ... & 2.19$^{+0.61}_{-1.18}$ & 2.22$^{+0.58}_{-0.85}$ & 2.24$^{+0.56}_{-0.85}$ & 2.21$^{+0.60}_{-0.88}$ & -1.53$^{+0.83}_{-0.39}$ & -1.46$^{+0.92}_{-0.44}$ & -1.46$^{+0.93}_{-0.45}$ & -1.14$^{+0.78}_{-0.65}$ \\
ln(E)    & 203.29 & ... & ... & ... & 203.39 & 203.36 & 203.13 & 202.95 & 203.26 & 203.18 & 202.66 & 202.90 \\
\hline
\textbf{WASP-77~A~b} & & & & & & & & & & & & \\
log(TiO) & ... & -6.65$^{+0.37}_{-0.26}$ & ... & -3.61$^{+0.57}_{-1.32}$ & ... & ... & ... & ... & ... & ... & ... & ... \\
log(VO)  & ... & ... & -8.93$^{+2.39}_{-2.17}$ & -9.02$^{+2.25}_{-2.19}$ & ... & ... & ... & ... & ... & ... & ... & ... \\
C/O      & 0.22$^{+0.32}_{-0.17}$ & 0.48$^{+0.29}_{-0.35}$ & 0.21$^{+0.29}_{-0.16}$ & 0.22$^{+0.30}_{-0.16}$ & ... & ... & ... & ... & ... & ... & ... & ... \\
log(Z)   & 2.49$^{+0.39}_{-3.65}$ & -0.96$^{+0.63}_{-0.37}$ & 2.56$^{+0.35}_{-0.82}$ & 2.59$^{+0.33}_{-0.55}$ & ... & ... & ... & ... & ... & ... & ... & ... \\
ln(E)    & 195.31 & 199.5 & 194.38 & 196.68 & ... & ... & ... & ... & ... & ... & ... & ... \\ 
\hline
\end{longtable}
\end{landscape}
}
\FloatBarrier

\begin{figure*}
   \centering
   \includegraphics[width=0.9\textwidth]{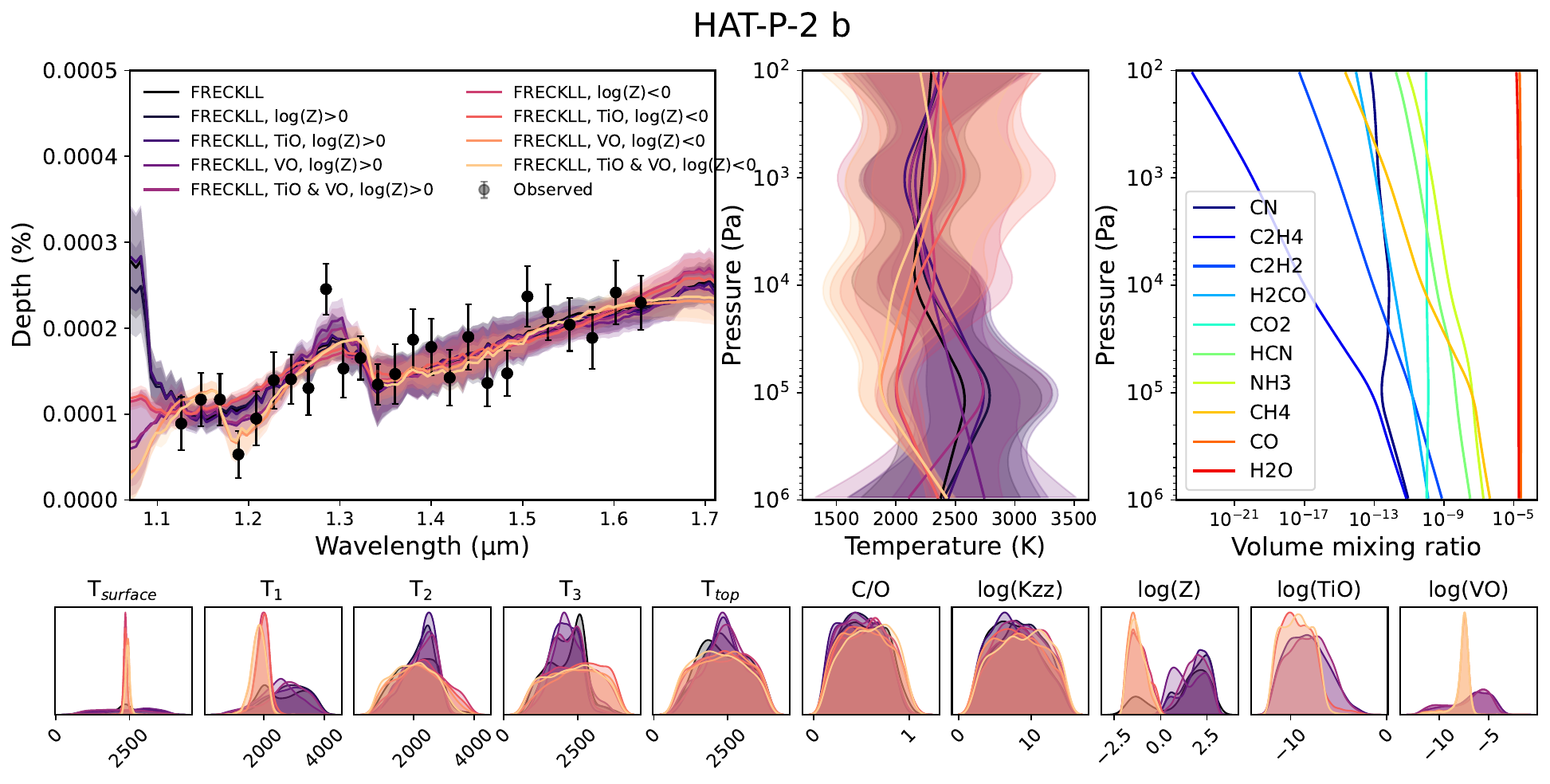}
   \caption{Detailed retrieval results for HAT-P-2 b: This planet is part of the group of planets for which the ``FRECKLL-only'' retrieval test depicted a bimodal distribution for the metalicity. Therefore, we have carried additional tests: we have tested the effect of FRECKLL, and the effect of FRECKLL plus TiO, VO and both TiO \& VO on two priors for the metallicity (one on the subsolar side log(Z)$<$0, and another one on the supersolar side log(Z)$>$0). 
   Upper row presents (from left to right) the fitting spectra for each retrieval configuration tested through this population study, the retrieved temperature profile, as well as the chemical structure of the planet only for the``FRECKLL-only'' retrieval.
   The lower row presents the posterior for each test, as normalised distribution.}
    \label{fig:HATP2b_indiv}
\end{figure*}

\begin{figure*}
    \centering
    \includegraphics[width=0.9\textwidth]{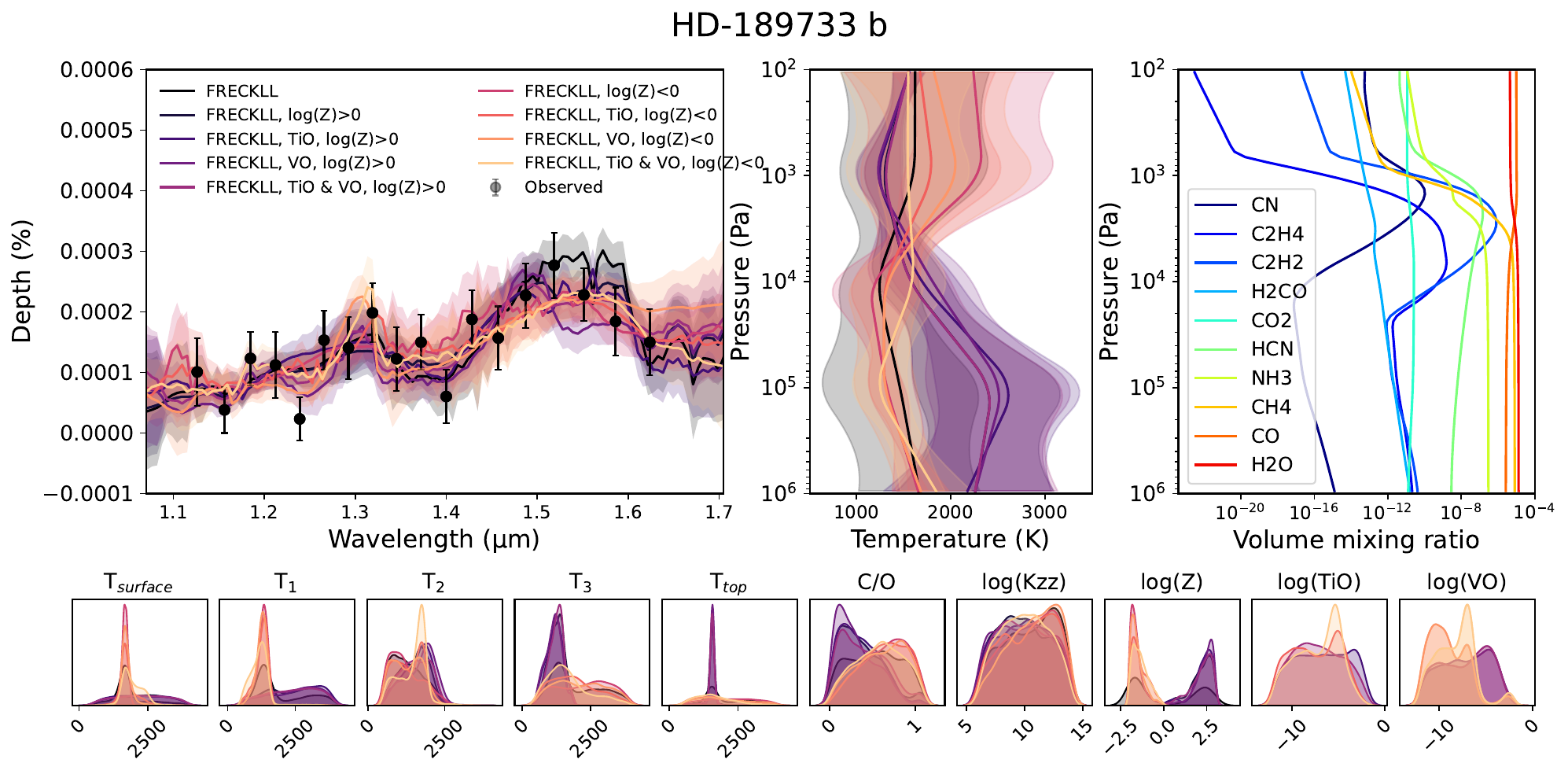}
    \caption{Detailed retrieval results for HD 189733 b: Same as Figure \ref{fig:HATP2b_indiv}, except that the chemical structure is shown for the retrieval ``FRECKLL, log(Z)$<$0'', as constraining the metallicity prior to the subsolar region improved ln(E) by 30.}
    \label{fig:HD189733b_emission}
\end{figure*}

\begin{figure*}
    \centering
    \includegraphics[width=0.9\textwidth]{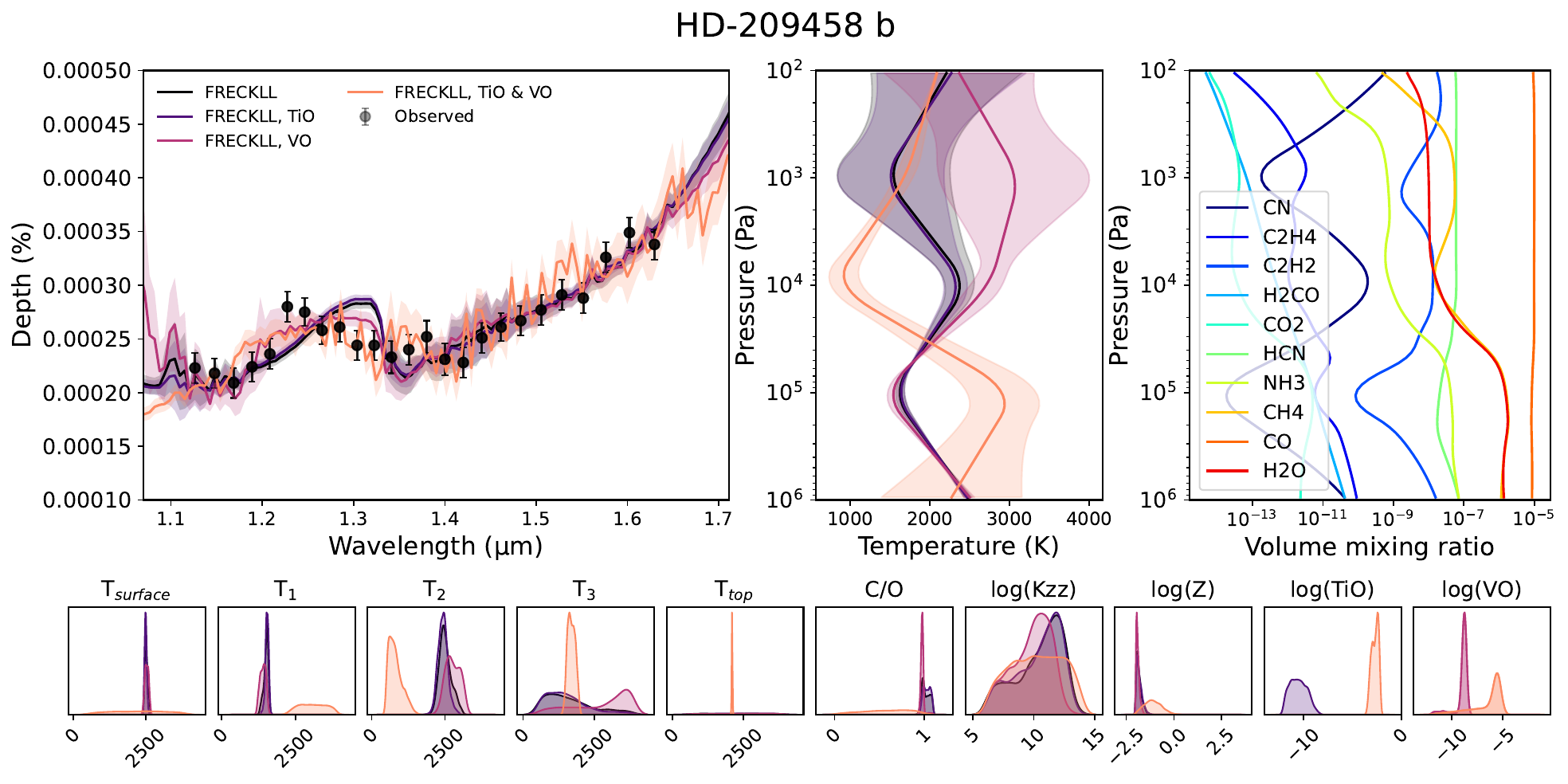}
    \caption{Detailed retrieval results for HD 209458 b: Upper row presents (from left to right) the fitting spectra for each retrieval configuration tested through this population study, the retrieved temperature profile, as well as the chemical structure of the planet only for the ``FRECKLL-only'' retrieval.
   The lower row presents the posterior for each test, as normalised distribution.}
    \label{fig:HD209459b_emission}
\end{figure*}

\begin{figure*}
   \centering
   \includegraphics[width=0.9\textwidth]{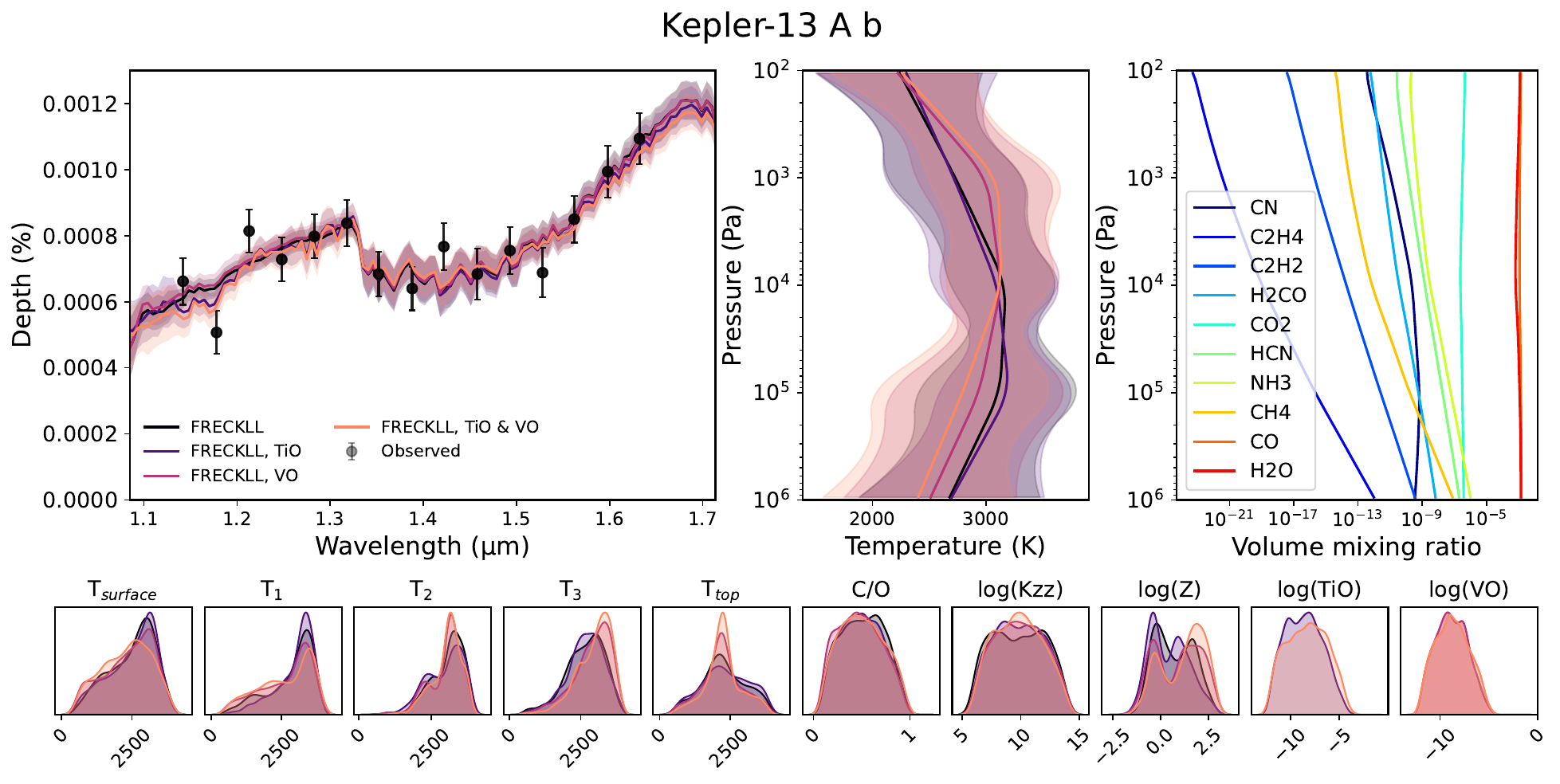}
   \caption{Same as Figure \ref{fig:HD209459b_emission} for Kepler-13 A b.}
    \label{fig:Kepler13Ab_indiv}
\end{figure*}

\begin{figure*}
   \centering
   \includegraphics[width=0.9\textwidth]{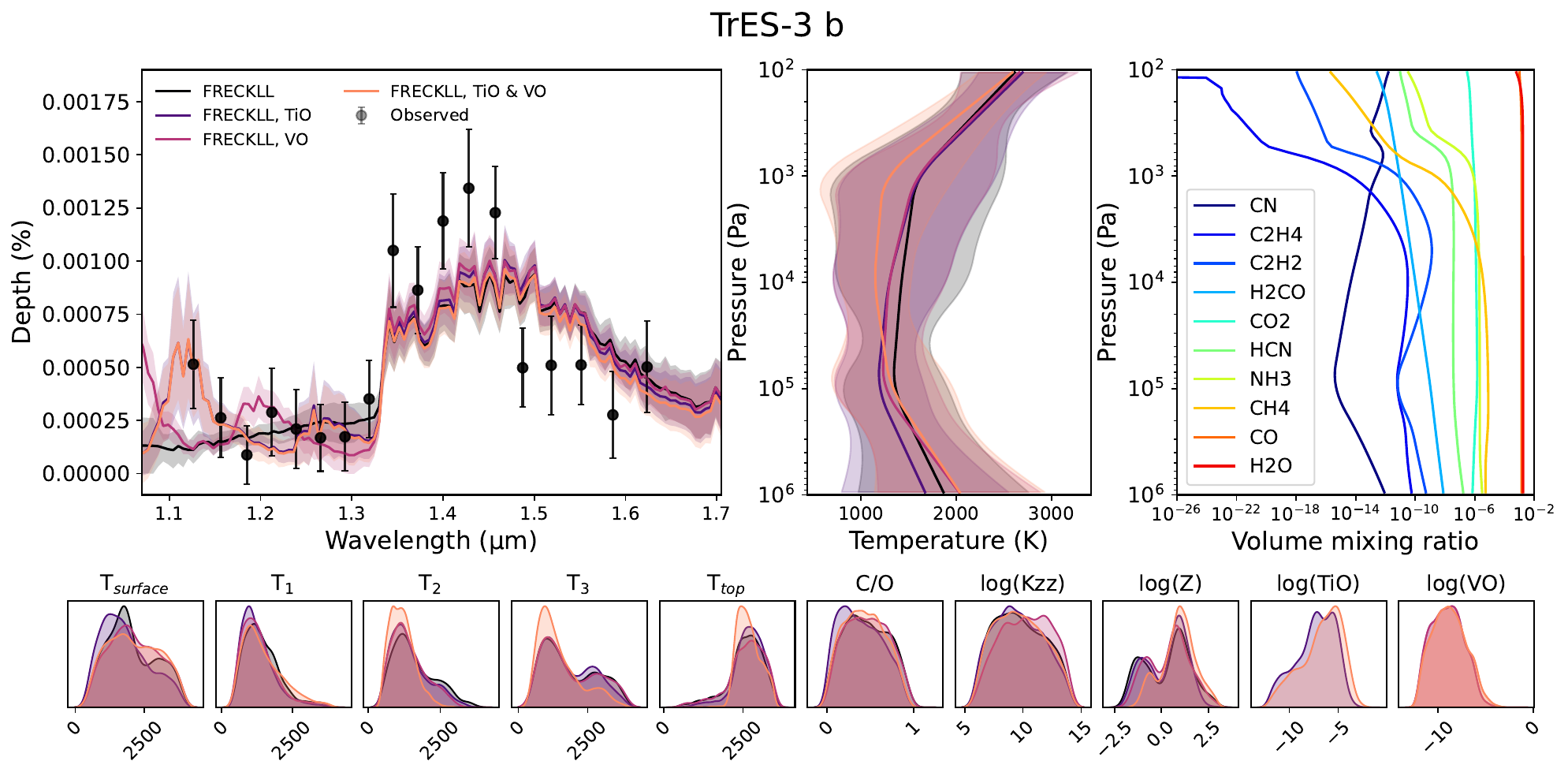}
   \caption{Same as Figure \ref{fig:HD209459b_emission} for TrES-3 b.}
    \label{fig:TrES3b_indiv}
\end{figure*}

\begin{figure*}
   \centering
   \includegraphics[width=0.9\textwidth]{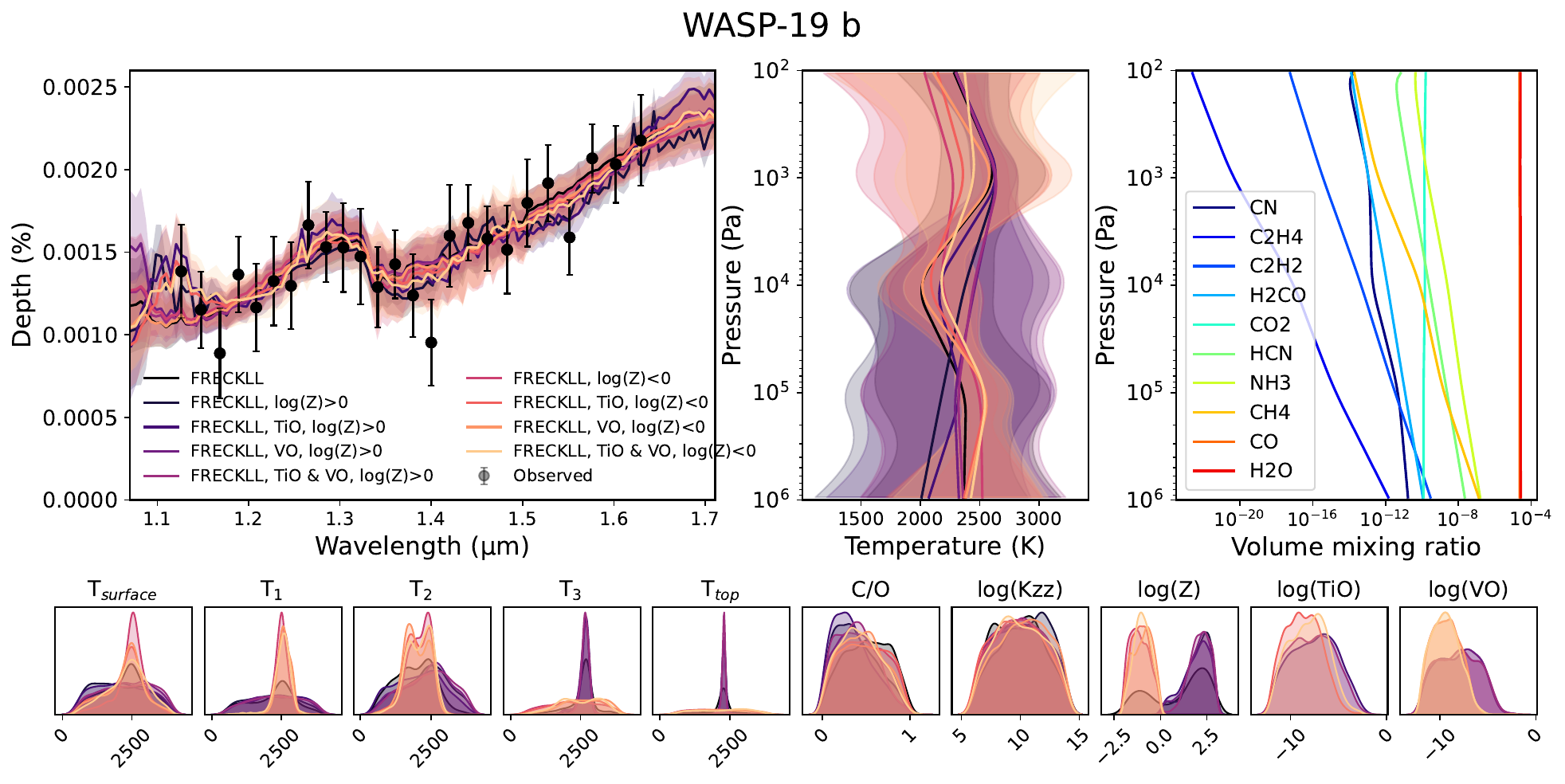}
   \caption{Same as Figure \ref{fig:HATP2b_indiv} for WASP-19 b.}
    \label{fig:WASP19b_indiv}
\end{figure*}

\begin{figure*}
   \centering
   \includegraphics[width=0.9\textwidth]{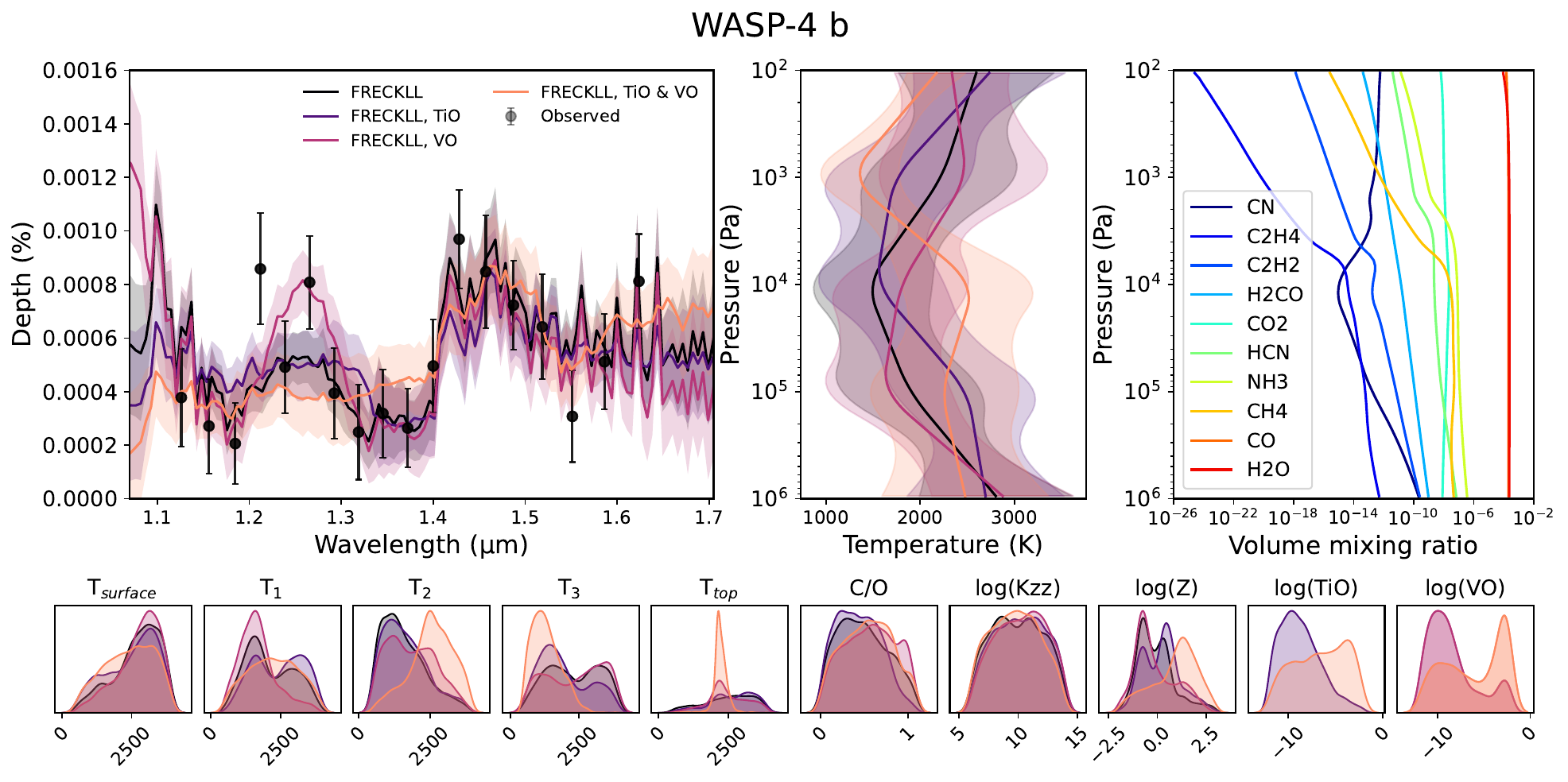}
   \caption{Same as Figure \ref{fig:HD209459b_emission} for WASP-4 b.}
    \label{fig:WASP4b_indiv}
\end{figure*}

\begin{figure*}
   \centering
   \includegraphics[width=0.9\textwidth]{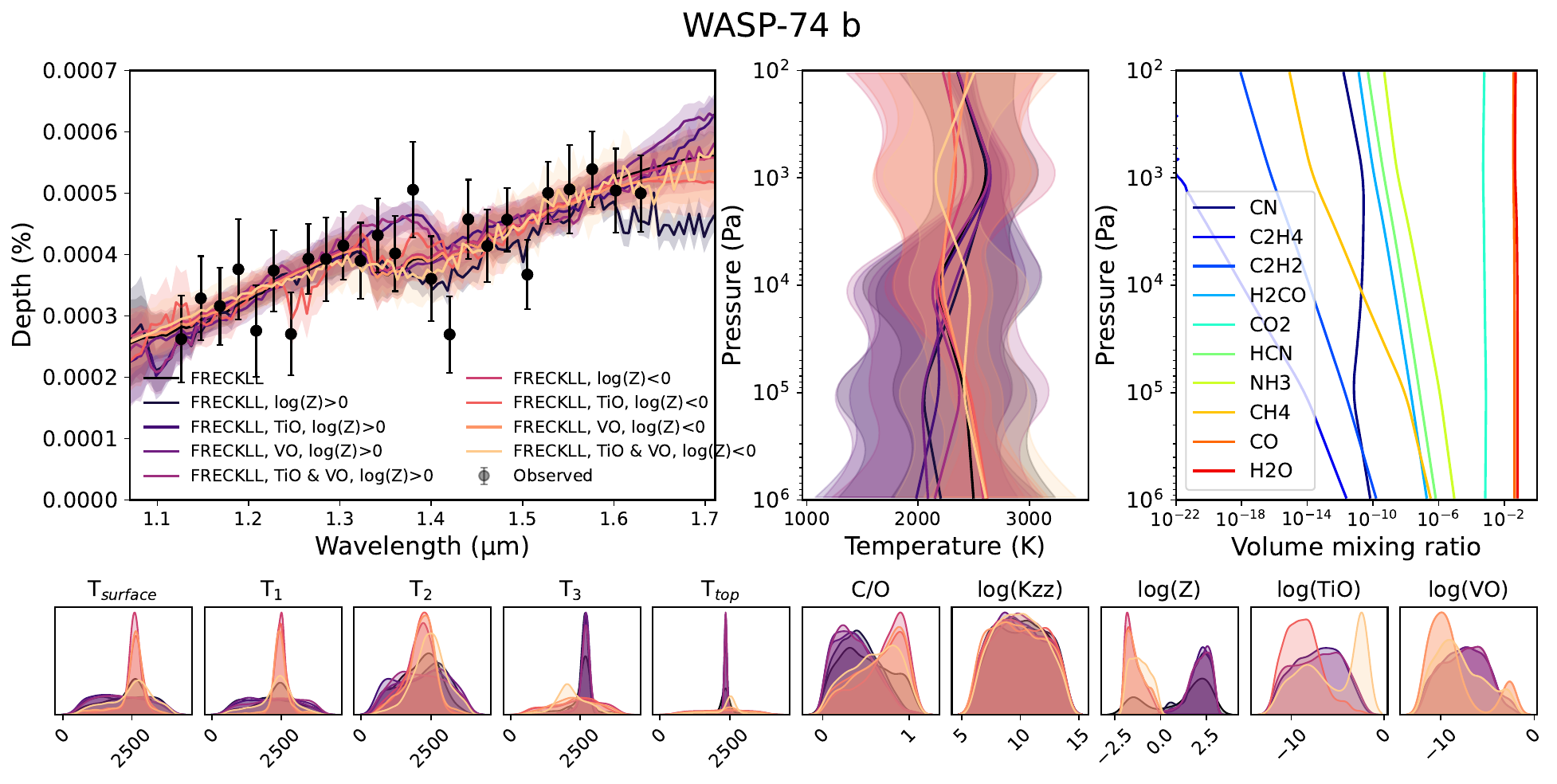}
   \caption{Same as Figure \ref{fig:HATP2b_indiv} for WASP-74 b.}
    \label{fig:WASP74b_emission}
\end{figure*}

\begin{figure*}
   \centering
   \includegraphics[width=0.9\textwidth]{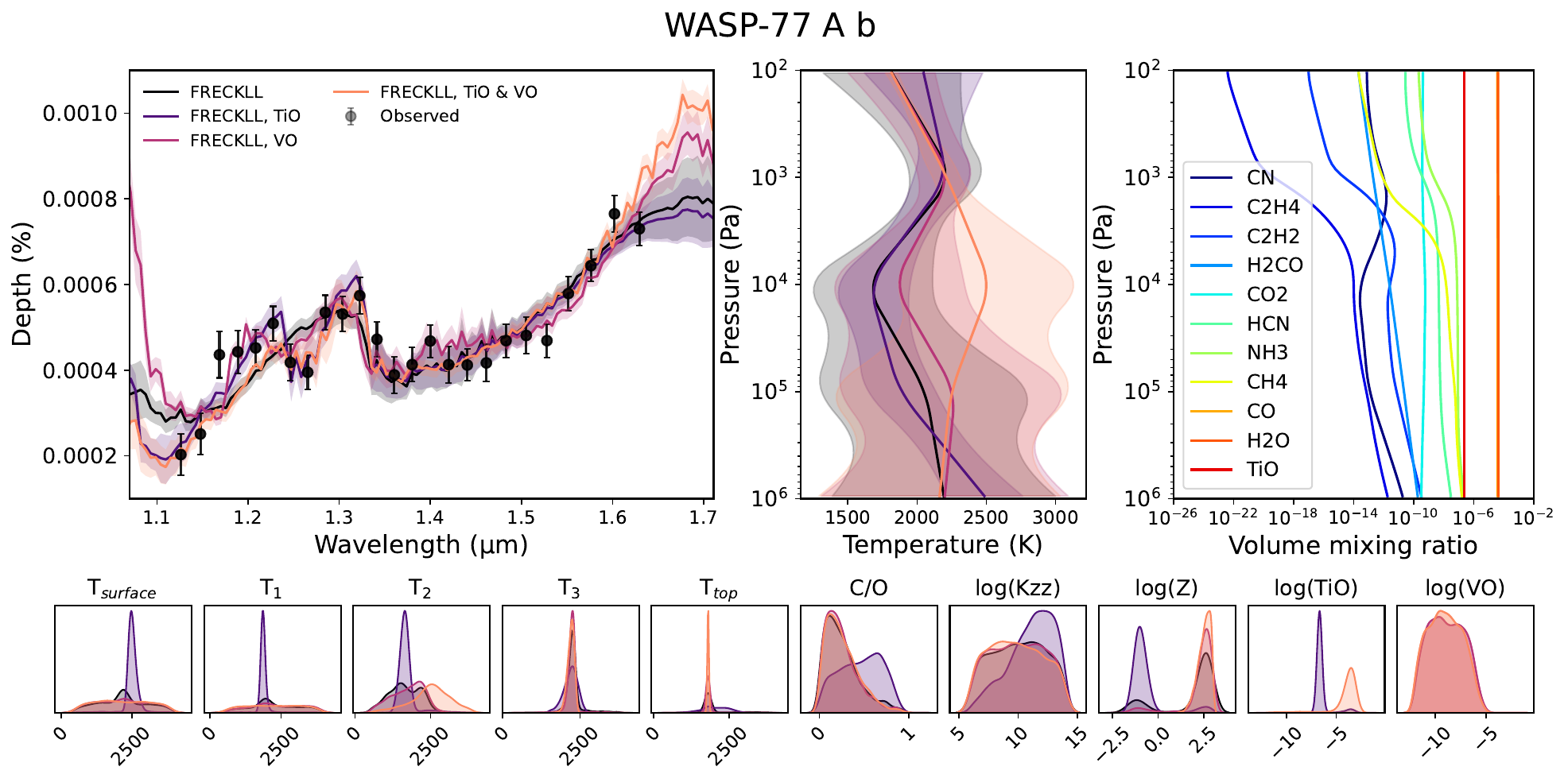}
   \caption{Same as Figure \ref{fig:HD209459b_emission} for WASP-77 A b, except the chemical profiles are resulting from the FRECKLL-TiO retrieval, which is the best fit obtained for this planet (see Table \ref{tab:eclipse_retrieval_results}).}
    \label{fig:WASP77Ab_indiv}
\end{figure*}

\FloatBarrier
\section{Individual planet analysis in transit}
\label{app:indiv_transit}

\begin{table}
\caption{\label{tab:transit_retrieval_results} Retrieval Results for the 4 Planets using Transit Spectra}
\centering
\begin{tabular}{lcc}
Parameters & FRECKLL & FRECKLL \\
           & 5-point thermal profile & isothermal profile \\
\hline
\hline
\textbf{HD 189733~b} & & \\
${\mathrm{R_{planet}}}$ & 1.197$^{+0.009}_{-0.001}$ & 1.209$^{+0.004}_{-0.008}$ \\
C/O      & 0.37$^{+0.38}_{-0.28}$ & 0.78$^{+0.26}_{-0.40}$ \\
log(Z)   & 0.16$^{+0.90}_{-1.61}$ & -0.58$^{+1.28}_{-1.14}$ \\
ln(E)    & 193.20 & 191.05 \\ 
\hline
\textbf{HD 209458~b} & & \\
${\mathrm{R_{planet}}}$ & 1.351$^{+0.017}_{-0.18}$ & 1.394$^{+0.002}_{-0.002}$ \\
C/O      & 0.61$^{+0.29}_{-0.39}$ & 0.21$^{+0.28}_{-0.15}$ \\
log(Z)   & 1.54$^{+0.36}_{-0.25}$ & -1.66$^{+0.60}_{-0.26}$ \\
ln(E)    & 204.88 & 208.35 \\ 
\hline
\textbf{WASP-43~b} & &  \\
${\mathrm{R_{planet}}}$ & 1.025$^{+0.004}_{-0.006}$ & 1.031$^{+0.001}_{-0.003}$ \\
C/O      & 0.42$^{+0.35}_{-0.24}$ & 0.62$^{+0.37}_{-0.38}$ \\
log(Z)   & 0.96$^{+1.33}_{-2.61}$ & -0.81$^{+3.43}_{-1.06}$ \\
ln(E)    & 196.82 & 195.63 \\ 
\hline
\textbf{WASP-74~b} & &  \\
${\mathrm{R_{planet}}}$ & 1.48$^{+0.02}_{-0.04}$ & 1.51$^{+0.01}_{-0.03}$ \\
C/O      & 0.61$^{+0.30}_{-0.35}$ & 0.48$^{+0.43}_{-0.30}$ \\
log(Z)   & 2.21$^{+0.56}_{-0.78}$ & 2.04$^{+0.77}_{-3.90}$ \\
ln(E)    & 194.24 & 194.50\\ 
\hline
\end{tabular}
\end{table}

\begin{figure*}
    \centering
    \includegraphics[width=0.9\textwidth]{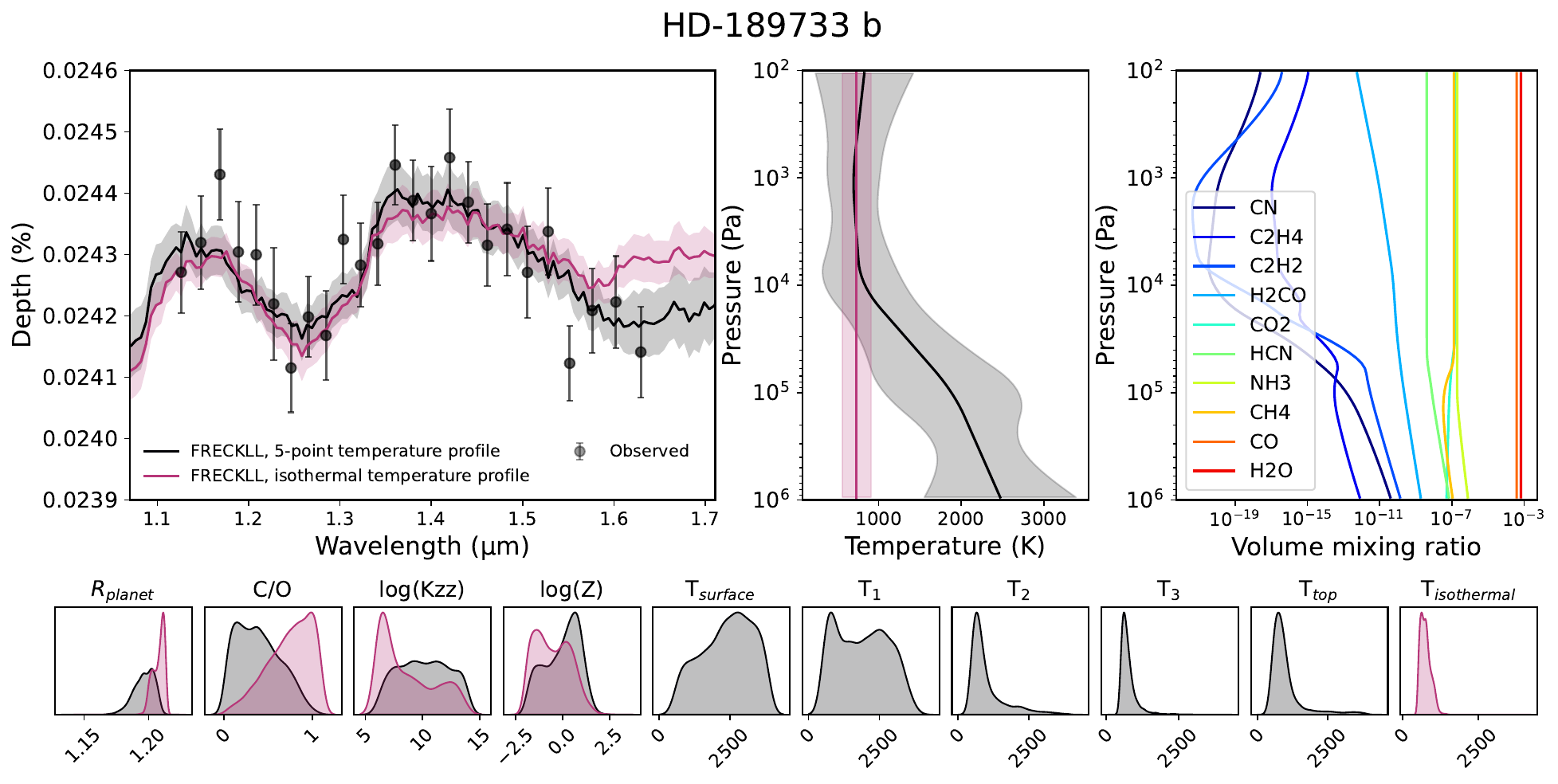}
    \caption{Detailed retrieval results for HD 189733 b: Upper row presents (from left to right) the fitting spectra for each retrieval configuration tested through this population study, the retrieved temperature profile, as well as the chemical structure of the planet only for the ``FRECKLL-5-point'' retrieval.
   The lower row presents the posterior for each test, as normalised distribution.}
    \label{fig:HD189733b_transmission}
\end{figure*}

\begin{figure*}
    \centering
    \includegraphics[width=0.9\textwidth]{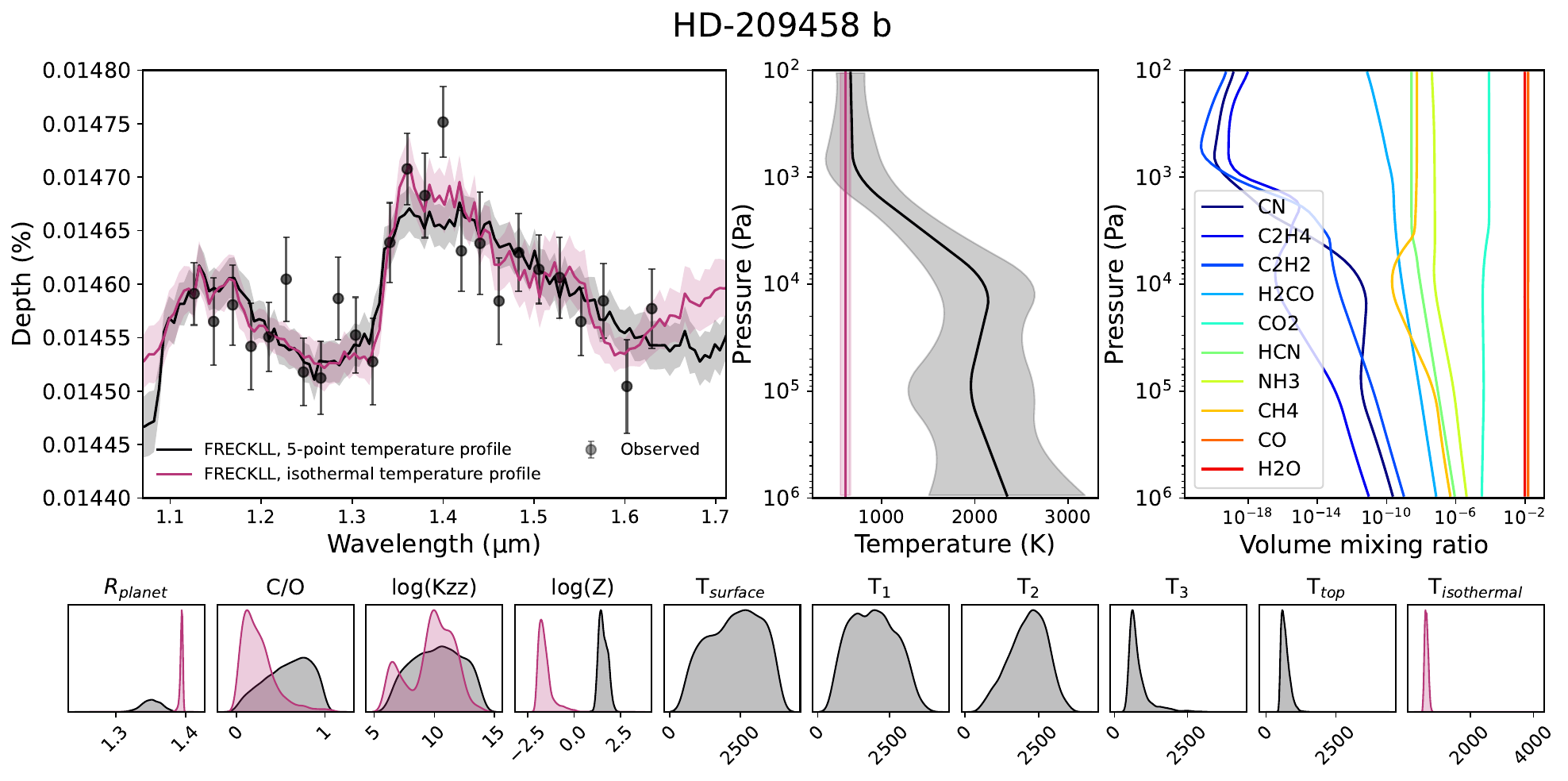}
    \caption{Same as Figure \ref{fig:HD189733b_transmission} for HD 209458 b.}
    \label{fig:HD209459b_transmission}
\end{figure*}

\begin{figure*}
   \centering
   \includegraphics[width=0.9\textwidth]{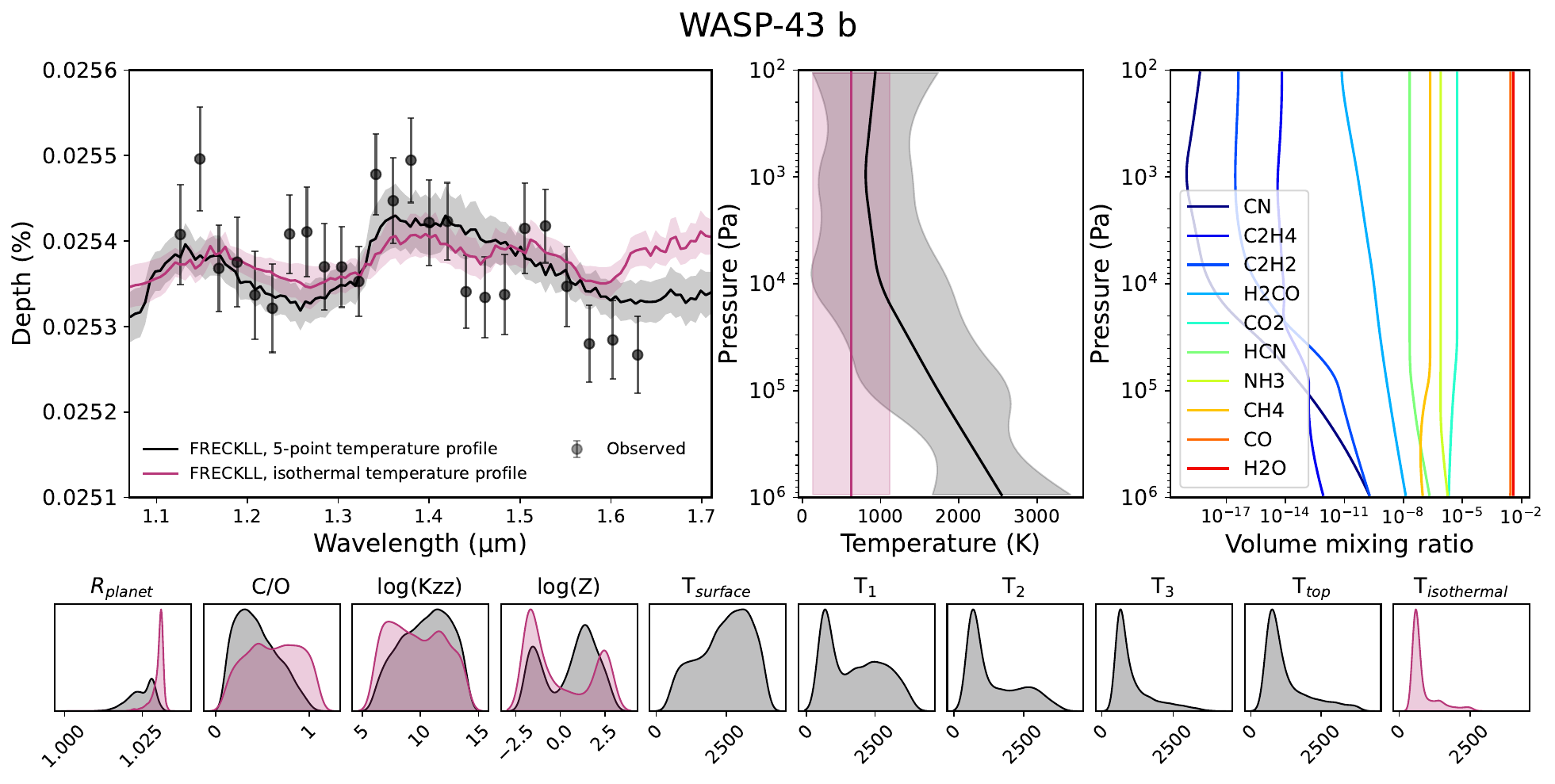}
   \caption{Same as Figure \ref{fig:HD189733b_transmission} for WASP-43 b.}
              \label{fig:WASP43b_transmission}
\end{figure*}

\begin{figure*}
   \centering
   \includegraphics[width=0.9\textwidth]{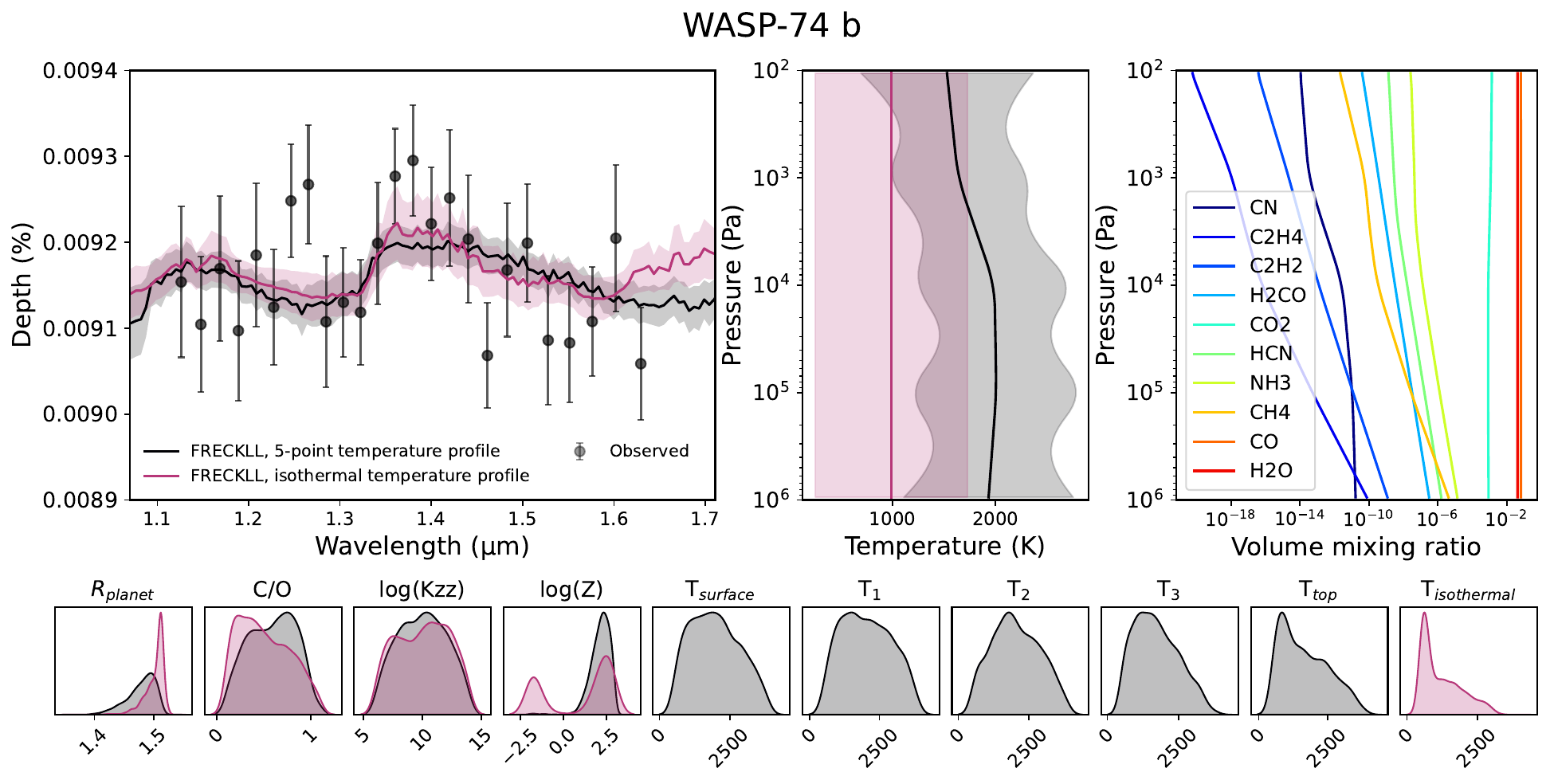}
   \caption{Same as Figure \ref{fig:HD189733b_transmission} for WASP-74 b.}
    \label{fig:WASP74b_transmission}
\end{figure*}

\end{appendix}

\end{document}